\documentclass[9pt]{extarticle}

\usepackage{geometry}
\usepackage[ruled]{algorithm2e}
\usepackage{appendix}
\usepackage{epsfig}
\usepackage{amsmath,amssymb,amsthm,amsfonts}
\usepackage{mathtools}
\usepackage{tikz}
\usepackage{color}
\usepackage{natbib}
\usepackage{graphicx}
\usepackage{subfigure}
\usepackage{caption}
\usepackage{url}
\usepackage{hyperref}
\usepackage{listings}
\usepackage{enumerate}
\usepackage{wrapfig}
\usepackage{pgfplots}
\usepackage{subfiles}
\usepackage{pbox}
\usepackage{csquotes}
\usepackage{float}
\numberwithin{equation}{section}

\newtheorem{theorem}{Theorem}

\newtheorem{example}{Example}

\graphicspath{{figures/}{../figures/}{figuresGaussian/}}

\usepackage{tikz}

\usetikzlibrary{shapes, arrows, calc, positioning,matrix}
\tikzset{
data/.style={circle, draw, text centered, minimum height=2.95em ,minimum width = .5em, inner sep = 2pt},
empty/.style={circle, text centered, minimum height=3em ,minimum width = .55em, inner sep = 2pt},
}

\tikzstyle{state}=[shape=circle,draw=black,fill=white, minimum height=3.0em ]
\tikzstyle{lightedge}=[<-,dotted]
\tikzstyle{mainstate}=[state,thick]
\tikzstyle{mainedge}=[<-,thick]
\tikzstyle{emptynode}=[shape=circle,draw=white,fill=white, minimum height=0.0em ]

\newcommand{\argmin}{\operatornamewithlimits{argmin}} 

\usepackage{setspace}
\begin{document}

  \vspace*{1.5cm}

 \begin{center}

{\noindent{\LARGE \textbf{ A generalised and fully Bayesian framework for ensemble updating } \vspace{1cm}\\} 
{\Large \textsc{Margrethe Kvale Loe}}\\{\it Department of
    Mathematical Sciences, Norwegian University of Science and
    Technology}\vspace{1cm}\\{\Large \textsc{H\aa kon Tjelmeland}}\\{\it Department of
    Mathematical Sciences, Norwegian University of Science and
    Technology}\vspace{1cm}
}	
\end{center}


\begin{abstract}
\sloppy
We propose a generalised framework for the updating of a prior ensemble to
a posterior ensemble, an essential yet challenging part in ensemble-based filtering methods. The proposed framework is based on a generalised and fully Bayesian
view on the traditional ensemble Kalman filter
(EnKF). In the EnKF,  the updating of the ensemble is based on Gaussian
assumptions, whereas in our general setup the updating may be based on another
parametric family. In addition, we propose to formulate an optimality
criterion and to find the optimal update with respect to
this criterion. 
The framework is fully Bayesian in the sense that  the  parameters of the assumed forecast model are treated as random variables. 
As a consequence, a  parameter vector is simulated, for each ensemble member, prior to the updating. 
 In contrast to existing fully Bayesian approaches, where the parameters are simulated conditionally on all the forecast samples, the parameters are in our framework simulated conditionally on both the data and all the forecast samples, except the forecast sample which is to be updated. 
 The proposed framework is studied in detail for two parametric
families. The first is for continuous variables, for which we use
the family of linear-Gaussian models and the optimality criterion is
to minimise the expected Mahalanobis distance between corresponding
prior and posterior ensemble members. For this situation, we find that the optimal
filter is a particular square root filter. The second parametric
family we study is the finite state-space hidden Markov model, where  the optimality criterion is to maximise the expected number of
elements in corresponding prior and posterior state vectors  that are equal.
For both cases,  we present simulation examples and compare the results
with existing ensemble-based filtering methods. The results of the proposed
approach indicate a promising performance. In particular, the filter based on the
linear-Gaussian model gives a more realistic representation of the
uncertainty than the traditional EnKF, and the effect of not conditioning on the forecast sample which is to be updated when simulating the parameters  is remarkable.

\end{abstract}


\vspace{0.5cm}
\noindent {\it } 
\vspace{-0.1cm}

\section{Introduction}
\label{sec:1}
 
The ensemble Kalman filter (EnKF) \citep{Burgers1998, art19} is a recursive Monte Carlo algorithm which provides an approximate solution to the filtering problem in statistics. 
The EnKF has been successfully applied to problems in several scientific fields, including reservoir modelling, oceanography and weather forecasting. Although the filter relies on a linear-Gaussian assumption about the underlying state-space model, it has shown to work well even in non-linear, non-Gaussian situations, and it also scales well to problems with very high-dimensional state vectors. The literature on the EnKF is extensive, and several modifications of the original algorithm of \cite{art2} have been proposed and studied. Much of the literature is quite geophysical-oriented with limited focus on the statistical foundations of the methodology. In recent years, however, the EnKF has gained increasing attention also from statisticians,  see for instance \cite{art22}. 
In the current report, we take a Bayesian perspective on the EnKF and use it to formulate a new and general class of ensemble filtering methods which also includes filtering of categorical variables.

The EnKF alternates between a forecast step and an update step. 
The main challenge, and the focus of this report, is the update step. The goal of the update step is to condition 
an ensemble of (approximate) realisations from a prior, or so-called forecast, distribution  on new observations   
so that a new ensemble of (approximate) realisations from the corresponding posterior, or so-called filtering, distribution is obtained. 
What causes trouble is that the forecast and filtering distributions are generally  intractable. 
To cope with this issue, the EnKF introduces Gaussian approximations and updates 
the forecast samples  in the form of a linear shift closely related to the linear update of the  mean in the traditional Kalman filter \citep{Kalman1960}. 
Since the resulting filtering ensemble is obtained from a linear shift of a possibly non-Gaussian forecast ensemble,  non-Gaussian  properties may have been captured.

An important feature about the linear update of the EnKF  is that it implicitly involves the construction of a Gaussian approximation to the forecast 
distribution. In practice, only a covariance matrix is  estimated. 
Combined with the assumption that the likelihood model is linear-Gaussian, the Gaussian approximation to the forecast distribution yields 
a Gaussian approximation to the filtering distribution according to  Bayes' rule. 
Under the assumption that the forecast ensemble contains independent samples from the Gaussian  approximation to the forecast model,  the linear shift corresponds to conditional simulation from a Gaussian distribution with mean and covariance so that each updated sample  marginally is distributed according to  the Gaussian approximation to the filtering distribution.  
\cite{LoeTjelmeland2020} present a generalisation of these underlying features of the EnKF and formulate a general class of ensemble updating procedures. 
The overall idea behind the framework they propose is that more generally another parametric model than the Gaussian can be pursued for the approximation to the forecast distribution. Likewise,  another parametric model than the linear-Gaussian can be pursued for the likelihood model. From Bayes' rule, a corresponding approximation  to the filtering distribution follows. 
To update the prior samples, the authors propose to simulate samples from a distribution conditional on the forecast ensemble such that, given that the forecast samples are distributed according to the constructed approximation to the forecast distribution, the updated samples are distributed according to the corresponding approximation to the filtering distribution,  which corresponds to the property of the EnKF linear update.

The traditional EnKF algorithm  is known to have a tendency to underestimate the variances in the forecast and filtering distributions, 
and the filter may in some cases even diverge. Various modifications have been proposed to correct for these issues, 
e.g.  localisation \citep{HoutekammerMitchellLocalisation2001, HamillWhitakerLocalisation2001, EdwardOttLocalEnsemble2004}  and inflation \citep{art26}. 
One possible reason for the unstable behaviour of the EnKF  is that uncertainty about the covariance matrix which is estimated from the forecast samples is not taken into account. 
That is, prior to the ensemble update, the covariance matrix of the Gaussian forecast approximation is estimated from the forecast ensemble, 
and thereafter the linear update proceeds as if this estimated covariance matrix were correct,  which obviously is not really the case even in a true linear-Gaussian situation. 
\cite{art20} address this issue and propose a Bayesian hierarchical EnKF (HEnKF) algorithm where  the mean and the covariance  
of the Gaussian forecast approximation are treated as random variables with prior distributions selected from the Gaussian conjugate family. 
Prior to the linear updating of the ensemble, the covariance matrix is then simulated rather than estimated. 
  \cite{art20} present simulation examples where their proposed HEnKF algorithm provides more reliable results than the traditional EnKF and reduces the undesirable effect of underestimating the variance. 
An improved version of the HEnKF algorithm is presented by \cite{Tsyrulnikov2017}. 
Other strategies for incorporating parameter uncertainty in the EnKF are proposed by \cite{art24} and  \cite{art23}. 
All studies indicate that it is advantageous to take parameter uncertainty into account. 

In the present report, we propose a fully Bayesian version of the framework proposed in  \cite{LoeTjelmeland2020}. The framework is fully Bayesian in the sense that the model parameters of the assumed forecast distri\-bution are treated as random variables. 
While the framework of \cite{LoeTjelmeland2020} can be seen as a gener\-alisation of the  traditional EnKF, the framework proposed in the present report can be seen as a gener\-alisation of the HEnKF of \cite{art20}, 
with one important modification.  
In \cite{art20}, a covariance matrix  is simulated for each ensemble member 
by simulating from the distribution of the covariance matrix given all the forecast samples. 
In a more general context, if we denote the parameters of the forecast model by $\theta$ and the forecast samples by $x^{(1)}, \dots, x^{(M)}$, where $M$ is the ensemble size, 
this would translate to simulating, for each ensemble member, a parameter vector $\theta^{(i)}$ from the distribution of $\theta$ given
$x^{(1)}, \dots, x^{(M)}$. 
In the present report, however, we propose to adopt a Bayesian model for the update from which it follows that 
also the incoming observation, say $y$, must be included in the conditioning, whilst the forecast sample $x^{(i)}$ to be updated must be excluded. 
In other words, prior to the updating of  $x^{(i)}$, we propose in this report to simulate a parameter $\theta^{(i)}$ conditionally on $y$ and $x^{(1)}, \dots, x^{(i-1)}, x^{(i+1)}, \dots, x^{(M)}$. 
Similarly to \cite{LoeTjelmeland2020}, we investigate two particular applications of the proposed framework: firstly, 
the case where the chosen forecast and likelihood approximations constitute a linear-Gaussian model, which corresponds to the model assumptions of the EnKF, and secondly, the case where the chosen forecast and likelihood approximations constitute a hidden Markov model (HMM) with categorical states. 
In contrast to \cite{LoeTjelmeland2020}, where  the  core focus is on the situation with the finite state-space HMM, this report also gives considerable focus 
to the linear-Gaussian model and the EnKF. In particular, we formulate a  class of EnKF algorithms, in a fully Bayesian setting, of which the traditional EnKF and the square root EnKF \citep{art5} represent special cases.

The remains of the report take the following outline. First, Section \ref{sec:2} provides some background material on state-space models and the EnKF. 
Next, our general ensemble updating framework is presented in Section \ref{sec:3}. In Sections \ref{sec:4} and \ref{sec:5} we consider 
two applications of the proposed framework, namely the linear-Gaussian model and the finite state-space HMM, respectively. 
In Sections \ref{sec:6} and \ref{sec:7},  we present simulation examples for the same two  cases. Finally, we finish off in Section \ref{sec:8} with a few closing remarks.

\section{Preliminaries}
\label{sec:2}

In this section, we describe state-space models and the related filtering problem in more detail. We also review the ensemble Kalman filter (EnKF). 

\subsection{State-space models}
\label{sec:2.1}

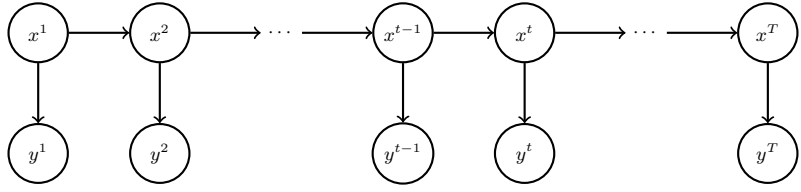
\begin{figure*}
    
    \centering
            \begin{tikzpicture}[thick,scale=0.8, every node/.style={scale=0.8}]
            [
                ->,
                >=stealth',
                auto,node distance=3cm,
                thick,
                main node/.style={circle, draw, font=\sffamily\Large\bfseries}
                ]
            
            \node[state] (x1) at (0,0) {$x^1$};
            \node[state] (x2) at (2,0) {$x^2$} edge [<-] (x1);
            \node[draw=none,fill=none] (ppp1) at (4,0) {$\cdots$} edge [<-] (x2);
            \node[state] (xt) at (6,0) {$x^{t-1}$} edge [<-] (ppp1);
            \node[state] (xt+1) at (8,0) {$x^{t}$} edge [<-] (xt);
            \node[draw=none,fill=none] (ppp2) at (10,0) {$\cdots$} edge [<-] (xt+1);
            \node[state] (xT) at (12,0) {$x^T$} edge [<-] (ppp2);
            
            \node[state] (y1) at (0,-2) {$y^1$} edge [<-] (x1);
            \node[state] (y2) at (2,-2) {$y^2$} edge [<-] (x2);
            \node[state] (yt) at (6,-2) {$y^{t-1}$} edge [<-] (xt);
            \node[state] (yt+1) at (8,-2) {$y^{t}$} edge [<-] (xt+1);
            \node[state] (yT) at (12,-2) {$y^T$} edge [<-] (xT);
            
            \end{tikzpicture}
    
    \caption{Graphical illustration of a general state-space model. }  
    \label{fig:SSM}
    \end{figure*}

A general state-space model consists of a latent process,  $\{x^t\}_{t=1}^{T}$, $x^t  = (x^t_1, \dots, x^t_n) \in \Omega_x$, 
and a corresponding observed process, $\{y^t\}_{t=1}^{T}$, $y^t = (y_1^t, \dots, y_m^t) \in \Omega_y$, with one observation $y^t$ for each  $x^t$.  
The latent $x^t$-process, usually called the state process,  constitutes a first order Markov chain with initial distribution $p(x^1)$ and transition probabilities 
$p_{x^t | x^{t-1}}(x^t | x^{t-1})$, $t\geq2$, so that the joint distribution of $x^{1:T} = (x^1, \dots, x^T)$ can be written as
$$
p_{x^{1:T}}(x^{1:T}) = p_{x^1}(x^1) \prod_{t=2}^T p_{x^t | x^{t-1}}(x^t | x^{t-1}).
$$
The observations $\{y^t\}_{t=1}^{T}$ are assumed conditionally independent given $\{x^t\}_{t=1}^{T}$, with  $y^t$ depending on $\{x^t\}_{t=1}^{T}$ only through $x^t$. 
Hence the joint likelihood for the observations $y^{1:T} = (y^1, \dots, y^T)$ can be written as
\[
p_{y^{1:T} | x^{1:T}}(y^{1:T} | x^{1:T} ) = \prod_{t=1}^T p_{y^t | x^t}(y^t | x^t).
\]
A graphical illustration of the general state-space model is shown in Figure \ref{fig:SSM}. 
When the variables of the state vector $x^t$ are categorical, the model is often called a hidden Markov model (HMM).  
Following \cite{chapter1},  the term HMM is in this report reserved for finite state-space state processes, 
while the term state-space model 
 may refer to either a categorical or a continuous situation. 

An important task associated with state-space models, and the main motivation for the work of this report, is the filtering problem. 
The objective of the filtering problem is, for each $t$, to compute the so-called filtering distribution, $p_{x^t | y^{1:t}}(x^t | y^{1:t})$, 
that is the distribution of the unobserved state $x^t$ given all the observations available at time $t$, $y^{1:t} = (y^1, \dots, y^t)$. 
Because of the particular state-space representation, the series of filtering distributions can be computed recursively according to a two-step procedure as follows:
\begin{equation}
p_{x^t | y^{1:t-1}}(x^t | y^{1:t-1}) = \int_{\Omega_x} p_{x^t | x^{t-1}}(x^t | x^{t-1}) p_{x^{t-1} | y^{1:t-1}}( x^{t-1} | y^{1:t-1}) \text{d} x^{t-1},
\label{eq:1}
\end{equation}
\begin{equation}
p_{x^t | y^{1:t}}(x^t | y^{1:t}) = \dfrac{p_{x^t | y^{1:t-1}}(x^t | y^{1:t-1}) p_{y^t | x^t}(y^t | x^t)}{\displaystyle \int_{\Omega_x} p_{x^t | y^{1:t-1}}(x^t | y^{1:t-1}) p_{y^t | x^t}(y^t | x^t) \text{d} x^t}.
\label{eq:2}
\end{equation}
The first step is called the prediction step and computes the forecast distribution $p_{x^t | y^{1:t-1}}(x^t | y^{1:t-1})$. 
The second step is called the update step and uses Bayes' rule to condition the forecast distribution 
on the incoming observation $y^t$ to compute the filtering distribution $p_{xt | y^{1:t}}(x^t | y^{1:t})$.  
The update step can be viewed as a standard Bayesian inference problem
where $p_{x^t | y^{1:t-1}}(x^t | y^{1:t-1})$ represents the prior, 
$p_{y^t | x^t}(y^t | x^t)$ the likelihood, and $p_{x^t | y^{1:t}}(x^t | y^{1:t})$ the posterior. 
For this reason,  the terms prior and forecast, and the terms posterior and filtering, are used interchangeably in this report. 

Although conceptually simple, the filtering recursions in Eqs. \eqref{eq:1} and \eqref{eq:2} are generally intractable because we are  unable to evaluate the integrals. 
Approximate solutions therefore become necessary. 
The most common approach is the class of simulation-based methods, or ensemble methods, where a set of samples, typically 
called an ensemble, is used to empirically represent the series of prediction and filtering distributions. 
Starting from an initial ensemble of independent realisations from the initial model $p_{x^1}(x^1)$, the idea is to 
advance the ensemble forward in time according to the state-space model dynamics. 
Similarly to the recursions in Eqs. \eqref{eq:1} and \eqref{eq:2}, ensemble methods alternate between a forecast step and an update step. 
Assuming at time $t$ that an ensemble $\{\tilde x^{t-1, (1)}, \dots, \tilde x^{t-1, (M)} \}$ 
of $M$ independent 
realisations from the previous filtering distribution $p_{x^{t-1} | y^{1:t-1}}(x^{t-1} | y^{1:t-1})$  is available, the forecast step is carried out by simulating 
$x^{t,(i)} | \tilde x^{t-1,(i)} \sim p_{x^t |  x^{t-1}}(x^t | \tilde x^{t-1, (i)})$ independently 
for each $i$. This yields a forecast ensemble, $\{ x^{t,(1)}, \dots, x^{t,(M)}\}$, with independent realisations from the forecast distribution $p_{x^t | y^{1:t-1}}(x^t | y^{1:t-1})$. Typically in practical applications, we are able to simulate from $p_{x^t | x^{t-1}}(x^t | x^{t-1})$, but often to a high computational cost, which restricts the ensemble size $M$ to be small. After the forecast step, the forecast ensemble needs to be updated taking the new observation $y^t$ into account, in order to obtain a new filtering ensemble, $\{\tilde x^{t, (1)}, \dots, \tilde x^{t, (M)} \}$, with independent realisations from the filtering distribution $p_{x^{t} | y^{1:t}}(x^{t} | y^{1:t})$ at time $t$. However, in contrast to the prediction step, there is no straightforward way to proceed with this updating. 
Therefore, ensemble filtering methods  re\-quire approximations in the update step. In the present report, we propose 
one such approximate updating method.

There exist two main classes of ensemble filtering methods: particle filters \citep{book1} and ensemble Kal\-man filters (EnKFs). Hybrid versions of these filters have also been proposed (e.g., \citealp{FreiKunsch2012, FreiKunsch2013}). In this report, we focus on the EnKF, and a brief review of the EnKF follows in the next section.

\subsection{The ensemble Kalman filter}
\label{sec:2.2}

The EnKF  is an ensemble filtering method which relies on Gaussian approximations in the update step. 
The filter was first introduced in \cite{art2} and several  modifications of the algorithm have been presented in the literature since then.
The variety of EnKF methods can be classified into two main categories, stochastic filters and deterministic filters, 
differing  in whether the updating of the ensemble is carried out in a stochastic or deterministic manner. Deterministic filters 
are also known as square root filters, and this is the term we use in this report. 

To understand the EnKF, consider first a linear-Gaussian model where 
$
x \sim  \mathcal N(x; \mu, Q)
$
and
$
y|x \sim \mathcal N(y; Hx, R),
$
 $\mu \in \mathbb R^{n}$, $Q \in \mathbb R^{n \times n}$, $H \in \mathbb R^{m \times n}$, and $R \in \mathbb R^{m \times m}$. 
The  posterior model corresponding to this linear-Gaussian model is  a Gaussian, $ \mathcal N(x; \mu^*, Q^*)$, with mean vector 
$\mu^* \in \mathbb R^{n}$ and covariance matrix $Q^* \in \mathbb R^{n \times n} $  analytically available from the Kalman filter equations as
\begin{eqnarray}
\mu^* = \mu + K(y - H \mu)
\label{eq:4}
\end{eqnarray}
and
\begin{eqnarray}
Q^* = (I_n-KH) Q, 
\label{eq:5}
\end{eqnarray}
respectively,
where $I_n \in \mathbb R^{n \times n}$ is the $n\times n$ identity matrix and 
\begin{equation}
K = Q H^\top \left  (H Q H^\top + R \right )^{-1}
\label{eq: Kalman}
\end{equation}
is the so-called Kalman gain matrix, where we have introduced  the notation $A^\top$  to denote the transpose of a matrix $A$. 
Now, suppose  $x \sim \mathcal N (x; \mu, Q)$ and $\epsilon \sim \mathcal N(\epsilon; 0, R)$ are independent random samples, 
and consider the linear transformation
\begin{equation}
\tilde x = x + K(y - Hx + \epsilon).
\label{eq:6}
\end{equation}
It is then a straightforward matter to show that    $\tilde x | y$ is distributed according to the Gaussian distribution $\mathcal N( x; \mu^*, Q^*)$ with mean  
$\mu^*$ and covariance $Q^*$  given by Eqs. \eqref{eq:4} and \eqref{eq:5}, respectively (e.g., \citealp{Burgers1998}). This result is used in the EnKF.

At a given time step $t$, the EnKF starts by making a linear-Gaussian assumption about the true (unknown) underlying model. 
Specifically, the forecast samples $x^{t,(1)}, \dots, x^{t,(M)}$ are assumed to be distributed according to a Gaussian distribution $\mathcal N(x^t ; \mu^t, Q^t)$ where the parameters $\mu^t$ and 
$Q^t$ are set equal to the sample mean and the sample covariance of the forecast ensemble, and  the likelihood model is assumed to be a Gaussian distribution with mean $H^tx^t$ and covariance $R^t$, $H^t \in \mathbb R^{m \times n}$, and $R^t \in \mathbb R^{m \times m}$. 
Under the assumption that the assumed  linear-Gaussian model is correct we have $x^{t,(i)} \sim N(x^t; \mu^t, Q^t)$ for each $i$, 
and the goal is to update $x^{t,(i)}$ so that $\tilde x^{t, (i)} \sim N(x^t; \mu^{*t}, Q^{*t})$, where $\mu^{*t}$ and $Q^{*t}$ are given by Eqs. \eqref{eq:4} and \eqref{eq:5}, respectively, with a superscript $t$ included in the notations, i.e. 
\begin{eqnarray}
\mu^{*t} = \mu^t + K^t(y^t - H^t \mu^t)
\label{eq:4..}
\end{eqnarray}
and
\begin{eqnarray}
Q^{*t} = (I_n-K^tH^t) Q^t, 
\label{eq:5..}
\end{eqnarray}
where, similarly, $K^t$ is given by Eq. \eqref{eq: Kalman}, with a superscript $t$ included,
$K^t  = Q^t (H^t)^\top \left  (H^t Q^t (H^t)^\top + R^t \right )^{-1}. $ 
The stochastic EnKF and the square root EnKF obtain this result in different ways.  
The stochastic EnKF proceeds by simulating $\epsilon^{t,(i)} \sim \mathcal N(\epsilon^t; 0, R^t)$ for $i=1, \dots, M$,  
and  then 
exploits  Eq. \eqref{eq:6}, which now takes the form 
\begin{equation}
\tilde x^{t,(i)} = x^{t,(i)} + K^t(y^t - H^tx^{t,(i)} + \epsilon^{t,(i)}).
\label{eq:stochEnKF}
\end{equation}
The square root EnKF takes a different approach and  instead performs a non-random linear transformation of $x^{t,(i)}$, 
\begin{equation}
\tilde x^{t,(i)} =   B^t(x^{t,(i)} - \mu^t)  + \mu^t +K^t(y^t - H^t\mu^t) ,
\label{eq:squareEnKF}
\end{equation}
where $B^t \in \mathbb R^{n \times n}$ is a solution to the quadratic matrix equation
\begin{equation}
B^t Q^t (B^t)^\top = (I_n-K^tH^t)Q^t.
\label{eq:7}
\end{equation}

If the underlying state-space model really is linear-Gaussian, the EnKF is consistent in the sense that the distribution of each  
updated sample converges to the true (Gaussian) filtering distribution as $M \to \infty$. In all other cases, the update is biased. 
However, since the posterior ensemble is obtained from a linear shift of a possibly non-Gaussian prior ensemble, 
non-Gaussian properties of the true prior and posterior models can, to some extent,  be captured.

\section{A general and fully Bayesian ensemble updating framework}
\label{sec:3}

In this section,  we formulate a general class of ensemble updating procedures.
As described in previous sections, the  goal is to update a given  ensemble of prior realisations, 
$ \{ x^{t,(1)}, \dots, x^{t,(M)}\}$, to a
corresponding ensemble of posterior realisations,  
$\{\tilde x^{t,(1)},\dots,\tilde x^{t,(M)}\}$, taking the new observation $y^t$
into account. To cope with this task, we propose to separately update each of the $x^{t,(i)}$ samples in
the prior ensemble to a corresponding $\tilde{x}^{t,(i)}$ sample in
the posterior ensemble, and to base the updating of
$x^{t, (i)}$ on an assumed Bayesian model.
As mentioned previously in the report, the proposed framework can be viewed as a generalisation of the 
hierarchical EnKF algorithm of \cite{art20} with the modification that the parameters are simulated in a different manner. 
The key steps of the proposed updating framework are summarised in Algorithm \ref{alg:1}. 

\begin{algorithm}[t]

\hspace{-0.325cm}1. Select the assumed distributions $f_{\theta^t}(\theta^t)$, $f_{x^t|\theta^t}({x^t|\theta^t})$ and $f_{y^t|x^t} (y^t|x^t)$ 
introduced in Section \ref{sec:3.1}

\hspace{-0.325cm}2.
 \For{$i = 1, \dots, M$}
 {
  \begin{itemize}
  
  \item[a)] Simulate $$\theta^{t,(i)} | x^{t,-(i)}, y^t \sim f_{\theta^{t} | x^{t,-(i)}, y^t} (\theta^{t} | x^{t,-(i)}, y^t)\; $$
  as described in Section \ref{sec:3.4}
  
  \item[b)] Construct the model $q(\tilde x^{t,(i)} | x^{t,(i)}, \theta^{t,(i)}, y^t)$ specified in Sections \ref{sec:3.2} and \ref{sec:3.3}
  
  \item[c)]   Simulate 
  $$
  \tilde x^{t,(i)} | x^{t,(i)}, \theta^{t,(i)}, y^t \sim q(\tilde x^{t,(i)} |  x^{t,(i)}, \theta^{t,(i)}, y^t)
  $$
  
  \end{itemize}
   }
 \caption{General ensemble updating procedure}
\label{alg:1}
\end{algorithm}

\subsection{Assumed Bayesian model}
\label{sec:3.1}

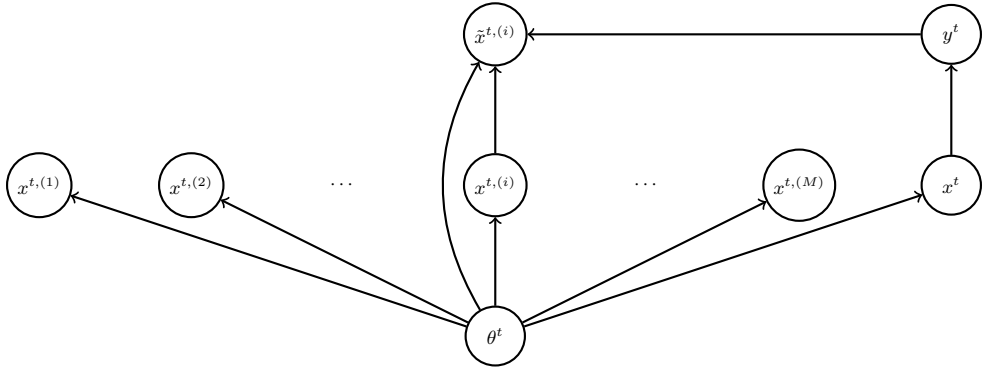
\begin{figure*}
    \centering

            \begin{tikzpicture}[thick,scale=0.8, every node/.style={scale=0.8}]
            [
                ->,
                >=stealth',
                auto,node distance=3cm,
                thick,
                main node/.style={circle, draw, font=\sffamily\Large\bfseries}
                ]
            
            \node[state] (theta) at (0, 0) {$\theta^t$};
            \node[state] (x1) at (-7.5,2.5) {$x^{t, (1)}$} edge [<-] (theta);
            \node[state] (x2) at (-5,2.5) {$x^{t, (2)}$} edge [<-] (theta);
            \node[draw=none, fill=none] (ppp) at (-2.5, 2.5) {$\cdots$};
            \node[state] (xi) at (0, 2.5) {$x^{t, (i)}$} edge [<-] (theta);
            \node[draw=none, fill=none] (ppp) at (2.5, 2.5) {$\cdots$};
            \node[state] (xM) at (5, 2.5) {$x^{t, (M)}$} edge [<-] (theta);
            \node[state] (x) at (7.5, 2.5) {$x^t$} edge [<-] (theta);
            \node[state] (y) at (7.5, 5) {$y^t$} edge [<-] (x);
            \node[state] (xiTilde) at (0, 5) {$\tilde x^{t, (i)}$} edge [<-] (xi) edge [<-] (y);
            \path[]
            (theta) edge [->,bend left] node [right] {} (xiTilde);
            
            \end{tikzpicture}
            
\caption{Graphical representation of the assumed Bayesian model for the updating of $x^{t, (i)}$ to $\tilde x^{t, (i)}$}

\label{fig: Bayesian model}
\end{figure*}

For the updating of the forecast sample $x^{t,(i)}$ we adopt an assumed Bayesian model. A graphical illustration of this assumed Bayesian model is shown in 
Figure \ref{fig: Bayesian model}. 
The model includes an unknown parameter vector 
$\theta^t \in \Omega_{\theta}$, and the forecast samples $x^{t,(1)},
\ldots, x^{t,(M)}$ and the latent state vector $x^t$ are assumed to be conditionally independent and
identically distributed given $\theta^t$. Moreover,  the
observation $y^t$ is assumed to be conditionally independent of $x^{t,(1)}, \ldots,
x^{t,(M)}$ and $\theta^t$ given $x^t$, and the updated sample 
$\tilde{x}^{t,(i)}$ is restricted to be conditionally independent of
$x^t$ and 
$$
x^{t,-(i)} = \{x^{t, (1)},\ldots,x^{t, (i-1)}, x^{t, (i+1)}, \dots, x^{t, (M)} \} 
$$
 given $x^{t, (i)}$, $\theta^t$ and $y^t$.

 To distinguish the assumed Bayesian model from the true and unknown underlying model, we use in the following the notation $f(\cdot)$ to denote distributions associated with the assumed Bayesian model, while, as in previous sections, $p(\cdot)$ is reserved for the truth. 
 Under the assumed Bayesian model,  the joint distribution of
$\theta^t$, $x^t$, $x^{t, (1)}, \dots, x^{t, (M)}$ and $y^t$ then reads
\begin{align*}
&f_{\theta^t, x^t, x^{t, (1)}, \dots, x^{t, (M)}, y^t}(\theta^t, x^t, x^{t, (1)}, \dots, x^{t, (M)}, y^t)
= 
f_{\theta^t}(\theta^t) f_{x^t|\theta^t}(x^t|\theta^t) f_{y^t|x^t}(y^t|x^t) \prod_{i=1}^M f_{x^t|\theta^t}(x^{t, (i)}|\theta^t),
\end{align*}
where  $f_{\theta^t}(\theta^t)$ is an assumed prior model for $\theta^t$, $f_{x^t|\theta^t}(x^t|\theta^t)$ is an
assumed prior model for $x^t|\theta^t$ and $f_{y^t|x^t}(y^t|x^t)$ is an assumed likelihood model. 
The prior $f_{x^t|\theta^t}({x^t|\theta^t})$ can be interpreted as an approximation to the intractable forecast model $p_{x^t | y^{1:t-1}}(x^t  | y^{1:t-1})$. 
The model $f_{\theta^t}(\theta^t)$ for $\theta^t$ should be chosen as a conjugate prior 
for $f_{x^t|\theta^t}(x^t|\theta^t)$, while the models $f_{x^t|\theta^t}(x^t|\theta^t)$ and
$f_{y^t|x^t}(y^t|x^t)$ must be chosen so that the corresponding posterior model
\[
f_{x^t|\theta^t, y^t}(x^t|\theta^t, y^t) \propto f_{x^t|\theta^t}(x^t|\theta^t) f_{y^t|x^t}(y^t|x^t) 
\]
is tractable.

\subsection{Class of updating distributions}
\label{sec:3.2}

Under the assumption that the assumed Bayesian model introduced above is correct, a na\"{i}ve updating procedure is to 
sample $\tilde x^{t,(i)}$ from $f_{x^t | x^{t,(1)}, \dots, x^{t,(M)}, y^t}(x^t | x^{t,(1)}, \dots, x^{t,(M)}, y^t)$. However, this procedure may be very sensitive to the assumptions of the assumed Bayesian model. 
To get an updating procedure which is more robust against the assumptions of the assumed model, a better approach is to generate $\tilde x^{t,(i)}$ as a modified version of $x^{t,(i)}$ and require
\begin{equation}
f_{\tilde x^{t, (i)} | x^{t, -(i)}, y^t} (x^t | x^{t, -(i)}, y^t)
=
f_{x^t | x^{t, -(i)}, y^t} (x^t| x^{t, -(i)}, y^t).
\label{eq:8}
\end{equation}
This way, we use the randomness in $x^{t,(i)}$ to generate randomness in $\tilde x^{t,(i)}$. The forecast sample $x^{t,(i)}$ is therefore not included in the conditioning in Eq. \eqref{eq:8}. 
To generate $\tilde x^{t,(i)}$ as a modified version of $x^{t,(i)}$ under this restriction, we introduce 
a distribution $q(\tilde x^{t,(i)} |  x^{t,(i)}, \theta^t, y^t )$ which fulfils Eq. \eqref{eq:8}, and simulate
$
\tilde x^{t,(i)} |  x^{t,(i)}, \theta^t, y^t  \sim q(\tilde x^{t,(i)} |  x^{t,(i)}, \theta^t, y^t ).
$
To construct such a $q(\tilde x^{t,(i)} |  x^{t,(i)}, \theta^t, y^t )$, we first note that the constraint in Eq. \eqref{eq:8} can be rewritten as
$$
\int_{\Omega_{\theta}}f_{\theta^t, \tilde x^{t,(i)} | x^{t, -(i)}, y^t}(\theta^t, x^t | x^{t,-(i)}, y^t) \text{d} \theta^t
= 
\int_{\Omega_{\theta}} f_{\theta^t, x^t | x^{t, -(i)}, y^t}(\theta^t, x^t | x^{t,-(i)}, y^t) \text{d} \theta^t . 
$$
Using that both $x^t$ and $\tilde x^{t, (i)}$ are conditionally independent of $x^{t, -(i)}$ given $\theta^t$ and $y^t$,
this  can be rewritten as
 \begin{equation}
\int_{\Omega_{\theta}} f_{\theta^t | x^{t, -(i)}, y^t}(\theta^t | x^{t, -(i)}, y^t) f_{\tilde  x^{t, (i)} |  y^t, \theta^t}( x^t |  y^t, \theta^t) \text{d} \theta^t
= \int_{\Omega_{\theta}} f_{\theta^t | x^{t, -(i)}, y^t}(\theta^t | x^{t, -(i)}, y) f_{x^t |  y^t, \theta^t}( x^t |  y^t, \theta^t) \text{d} \theta^t. 
 \label{eq:9}
 \end{equation}
A sufficient condition for this restriction to hold  is
\begin{equation}
f_{\tilde  x^{t, (i)} |   \theta^t, y^t}( x^t |  \theta^t, y^t) = f_{x^t |   \theta^t, y^t} (x^t |   \theta^t, y^t)
\label{eq:10}
\end{equation}
for all $x^t, \theta^t$, and $y^t$. 
Thereby,  if for a given $\theta^t$  we  can manage to construct a $q(\tilde x^{t, (i)} | x^{t, (i)}, \theta^t, y^t)$ consistent 
with Eq. \eqref{eq:10}, we can update  $x^{t, (i)}$  by first simulating 
$\theta^{t, (i)} | x^{t, -(i)}, y^t \sim f_{\theta^t | x^{t, -(i)}, y^t}(\theta^t |
x^{t, -(i)}, y^t)$ and thereafter simulate  
$\tilde x^{t, (i)} | x^{t, (i)}, \theta^{t, (i)}, y^t \sim q(\tilde x^{t, (i)} | x^{t, (i)}, \theta^{t, (i)}, y^t)$. 
How to simulate $\theta^{t, (i)} | x^{t, -(i)}, y^t$ 
is discussed  in Section \ref{sec:3.4}. 
To construct a $q(\tilde x^{t, (i)} | x^{t, (i)}, \theta^{t, (i)}, y^t)$ consistent with 
Eq. \eqref{eq:10} we note that for $f_{\tilde  x^{t, (i)} |  \theta^t, y^t}( x^t |   \theta^t, y^t)$ in Eq. \eqref{eq:10} we have 
\begin{align*}
f_{\tilde  x^{t, (i)} |  \theta^t, y^t}( \tilde x^{t, (i)} |  \theta^t, y^t) =  \int_{\Omega_x} f_{x^t|\theta^t}(x^{t}|\theta^t) q(\tilde x^{t, (i)} | x^{t}, \theta^t, y^t) \text{d} x^{t}. 
\end{align*}
Thereby, from Eq. \eqref{eq:10}, it follows that $q(\tilde x^{t, (i)} | x^{t, (i)}, \theta^t, y^t)$ must fulfil 
\begin{align}
f_{x^t |  \theta^t, y^t} (\tilde x^{t, (i)} |  \theta^t, y^t) =  \int_{\Omega_x} f_{x^t|\theta^t}(x^{t}|\theta^t) q(\tilde x^{t,(i)} | x^{t}, \theta^t, y^t) \text{d} x^{t}
\label{eq:11}
\end{align}
for all $\tilde x^{t, (i)}, \theta^t$ and $y^t$.

The criterion in Eq.  \eqref{eq:11} defines a class of updating distributions 
 in the sense that there  may be infinitely many solutions $q(\tilde x^{t, (i)} | x^{t, (i)}, \theta^t, y^t)$ 
 which fulfil Eq. \eqref{eq:11}.   It should be noted that if the assumed model  is correct it 
 does not matter which  $q(\tilde x^{t, (i)} | x^{t, (i)}, \theta^t, y^t)$ within this class we choose; 
the distribution of $\tilde x^{t, (i)} | x^{t, -(i)}, y^t$ then equals $f_{x^t|x^{t, -(i)}, y^t}(x^t | x^{t, -(i)}, y^t)$ regardless. 
Generally, however, the assumed model is wrong, 
and the choice of $q(\tilde x^{t, (i)} | x^{t, (i)}, \theta^t, y^t)$ can have a substantial effect on the actual 
distribution of $\tilde x^{t, (i)} | x^{t, -(i)}, y^t$.  The simplest
solution is to set $ q(\tilde x^{t, (i)} | x^{t, (i)}, \theta^t, y^t)$ equal 
to $f_{x^t |  \theta^t, y^t} (x^t |  \theta^t, y^t)$ which entails  that we simulate $\tilde x^{t, (i)}$ independently 
of $x^{t, (i)}$. However, this na\"{i}ve approach is very sensitive to the assumptions of the assumed model and is not a good way to proceed as  
we loose a lot of valuable information from $x^{t, (i)}$ about the {true} (unknown) model 
that we may not have been able to capture with the assumed model. As discussed above, we want to generate 
$\tilde x^{t, (i)}$  as a modified version of $x^{t, (i)}$.  That way, we retain more information from $x^{t,(i)}$ about the true model. 
An optimal solution $ q(\tilde x^{t, (i)} | x^{t, (i)}, \theta^t, y^t)$  within the class of distributions can be found if an optimality criterion is specified, 
which we discuss in the next section.

\subsection{Optimality criterion}
\label{sec:3.3}

Generally, an optimal solution, denoted $q^*(\tilde x^{t, (i)} | x^{t, (i)}, \theta^t, y^t)$, with\-in the class of distributions 
defined in the previous section can for example be defined as the solution which minimises the expected value of some 
function $g(x^{t, (i)}, \tilde x^{t, (i)})$, 
\[
q^*(\tilde x^{t, (i)} | x^{t, (i)}, \theta^t, y^t) = \argmin_{q(\cdot)} \text{E} \left [ g(x^{t, (i)}, \tilde x^{t, (i)})  \right ],
\]
where the expectation is taken over the distribution $f_{x^t|\theta^t}(x^{t, (i)}|\theta^t) q(\tilde x^{t, (i)} | x^{t, (i)}, \theta^t, y^t)$,
i.e. the joint distribution of $x^{t, (i)}$ and $\tilde x^{t, (i)}$ given $(\theta^t,y^t)$ under the assumption that the assumed Bayesian model is correct. 
In the present report,  we propose to choose the  function  $g(x^{t,(i)}, \tilde x^{t,(i)})$  as the Mahalanobis distance between $x^{t,(i)}$ and $\tilde x^{t,(i)}$, 
\begin{equation}
g(x^{t,(i)}, \tilde x^{t,(i)}) =  \left ( x^{t,(i)} - \tilde x^{t,(i)} \right )^\top \Sigma^{-1}  \left ( x^{t,(i)} - \tilde x^{t,(i)} \right ), 
\label{eq:12}
\end{equation}
where $\Sigma \in \mathbb R^{n \times n}$ is some positive definite matrix. 
If $\Sigma$ equals the identity matrix,   $g(x^{t,(i)}, \tilde x^{t,(i)})$  reduces to the squared Euclidean distance between $x^{t,(i)}$ and $\tilde x^{t,(i)}$,
\begin{equation}
g(x^{t,(i)}, \tilde x^{t,(i)}) = \sum_{j=1}^n \left (x_j^{t,(i)} - \tilde x_j^{t,(i)} \right )^2.
\label{eq:13}
\end{equation}
Basically, the optimality criterion then states  that we want to make minimal changes to each prior 
sample $x^{t, (i)}$. To us, this seems like a reasonable criterion since we want to capture as much information from  
$x^{t, (i)}$ as possible. Of course, one must value the information that comes with the observation
$y^t$, but there is no reason to make more changes to $x^{t, (i)}$ than necessary. 

If  $x^t$ is a vector of categorical variables, $x_j^t \in \{0, 1, \dots, K-1\}$, an alternative is to select 
$g(x^{t, (i)}, \tilde x^{t, (i)})$ as the number of corresponding elements of $x^{t, (i)}$ and $\tilde x^{t, (i)}$ that are different, 
\begin{equation}
g(x^{t, (i)}, \tilde x^{t, (i)}) = \sum_{j=1}^n 1\left (x_j^{t, (i)} \neq \tilde x_j^{t, (i)} \right ),
\label{eq:14}
\end{equation}
where $1(\cdot)$ denotes the usual indicator function. If each component $x_j^t$ is {binary}, 
the functions in Eqs. \eqref{eq:13} and \eqref{eq:14} are equal.

\subsection{Parameter simulation }
\label{sec:3.4}

In this section, we describe  how to simulate  $\theta^{t, (i)} | x^{t, -(i)}, y^t \sim f_{\theta^t|x^{t, -(i)}, y}(\theta^t | x^{t, -(i)}, y^t)$ when   
$f_{\theta^t}(\theta^t)$ is chosen as a conjugate prior for $f_{x^t|\theta^t}(x^t|\theta^t)$. 
Specifically, we can then introduce $x^t$ as an auxiliary variable and simulate $(x^t, \theta^t)$ from the 
joint distribution
\begin{align*}
f_{x^t, \theta^t|x^{t, -(i)},y^t}(x^t, \theta^t | x^{t, -(i)}, y^t) \propto f_{\theta^t}(\theta^t) f_{x^t|\theta^t}(x^t | \theta^t) f_{y^t|x^t}(y^t|x^t) \prod_{j\neq i} f_{x^t|\theta^t}(x^{t, (j)} | \theta^t)
\end{align*}
by constructing  a Gibbs sampler which alternates between drawing $x^t$
from the full conditional distribution $f_{x^t|\theta^t, x^{t, -(i)},y^t}(x^t | \theta^t, x^{t, -(i)}, y^t)$ and $\theta^t$ from the full conditional distribution 
$f_{\theta^t | x^t, x^{t, -(i)}, y^t}(\theta^t | x^t, x^{t, -(i)}, y^t)$.  
Using that $x^t$ and $x^{t,-(i)}$ are conditionally independent given $\theta^t$  (see Figure \ref{fig: Bayesian model}), it follows  that the full conditional distribution  $f_{x^t|\theta^t,x^{t, -(i)},y^t}(x^t | \theta^t, x^{t, -(i)}, y^t)$ is given as  
\[
f_{x^t | \theta^t, x^{t, -(i)}, y^t}(x^t | \theta^t, x^{t, -(i)}, y^t) = f_{x^t| \theta^t, y^t}(x^t |  \theta^t, y^t).
\]
Simulating from $ f_{x^t| \theta^t, y^t}(x^t |  \theta^t, y^t)$ should be achievable,  since  
$f_{x^t|\theta^t}(x^t|\theta^t)$ and $f_{y^t|x^t}(y^t|x^t)$ are chosen so that $f_{x^t| \theta^t, y^t}(x^t |  \theta^t, y^t)$ is 
tractable. 
Using that $\theta^t$ and $y^t$ are conditionally independent given $x^t$ (again, see Figure \ref{fig: Bayesian model}),  the other  full conditional distribution,  $f_{\theta^t | x^t, x^{t, -(i)}, y^t}(\theta^t | x^t, x^{t, -(i)}, y^t)$, is given as 
\[
f_{\theta^t | x^t, x^{t, -(i)}, y^t}(\theta^t | x^t, x^{t, -(i)}, y^t) = f_{\theta^t |x^t, x^{t, -(i)}}(\theta^t | x^t, x^{t, -(i)}).
\]
Since $f_{\theta^t}(\theta^t)$ is chosen as a conjugate prior for $f_{x^t | \theta^t}(x^t | \theta^t)$, and since $x^t$, $x^{t,(1)}, \dots, x^{t,(M)}$ are independent and identically distributed given $\theta^t$, it follows that  $f_{\theta^t | x^t, x^{t, -(i)}}(\theta^t | x^t, x^{t, -(i)})$  is tractable and 
belongs to the same family of distributions as $f_{\theta^t}(\theta^t)$. Simulating from  $f_{\theta^t | x^t, x^{t, -(i)}}(\theta^t | x^t,
x^{t, -(i)})$ should thereby also be possible.

\section{Application 1: Linear-Gaussian assumed model}
\label{sec:4}

In this section, we consider how the updating procedure described in  Section \ref{sec:3} can be applied when the elements of the state vector $x^t$ are  continuous variables. Specifically, we then propose to let the distributions $f_{x^t | \theta^t}(x^t | \theta^t)$ and $f_{y^t | x^t}(y^t | x^t)$ of the assumed Bayesian model 
constitute a  linear-Gaussian model. 
As we shall see, the resulting optimal updating procedure then corresponds to a fully Bayesian version of a square root EnKF.

\subsection{Specification of the assumed model }
\label{sec:4.1}
Suppose $x^t = (x_1^t, \dots, x^t_n) \in \mathbb R^n$ and $y^t = (y^t_1, \dots, y^t_m) \in \mathbb R^m$. Let 
$\theta^t = ( \mu^t, Q^t)$ where $\mu^t \in \mathbb R^n$, $Q^t \in \mathbb R^{n \times n}$, and $Q^t$ is positive definite. Select  $f_{x^t|\theta^t}(x^t|\theta^t)$ as a 
Gaussian distribution with mean vector $\mu^t$ and covariance matrix $Q^t$, 
\[
f_{x^t|\theta^t} (x^t|\theta^t) = \mathcal N(x^t; \mu^t, Q^t),
\]
and choose $f_{y^t|x^t}(y^t|x^t)$ as a Gaussian distribution with mean $H^tx^t$, $H^t \in \mathbb R^{m \times n}$ and covariance matrix $R^t \in \mathbb R^{m \times m}$,
\[
f_{y^t|x^t}(y^t|x^t) = \mathcal N(y^t; H^tx^t, R^t). 
\]
Given $\theta^t$, this model corresponds to the linear-Gaussian model introduced in Section \ref{sec:2.2}. The corresponding posterior model   
$f_{x^t|\theta^t,y^t}(x^t|\theta^t, y^t)$ is then a Gaussian distribution with mean vector $\mu^{*t}$ and covariance matrix $Q^{*t}$ given by Eqs. \eqref{eq:4..} and \eqref{eq:5..}, respectively.
Following Section \ref{sec:3}, we adopt a conjugate prior for $\theta^t$, which in this case entails an inverse Wishart distribution for  $Q^t$,
\begin{equation}
\label{eq:15}
f_{Q^t}(Q^t) = \mathcal W^{-1}(Q^t; V, \nu),
\end{equation}
and a Gaussian distribution for $\mu^t | Q^t$,
\begin{equation}
\label{eq:16}
 f_{\mu^t | Q^t} (\mu^t | Q^t) = \mathcal N(\mu^t; \mu_0, \kappa^{-1} Q^t ),
\end{equation}
 where $\nu, \kappa \in \mathbb R$, $\mu_0 \in \mathbb R^{n}$ and $V \in \mathbb R^{n \times n}$ are known hyperparameters.

\subsection{Derivation of the class of updating distributions}
\label{sec:4.2}

The restriction in Eq.  \eqref{eq:10} now entails that the updating distribution $q(\tilde x^{t, (i)} | x^{t, (i)}, \theta^t, y^t) $ 
must be chosen so that the integral on the right hand side of  Eq. \eqref{eq:11} returns a Gaussian distribution 
with mean vector equal to $\mu^{*t}$ in Eq. \eqref{eq:4..} and covariance matrix equal to $Q^{*t}$ in Eq. \eqref{eq:5..}. To obtain this, we start by 
selecting $ q (\tilde x^{t, (i)} | x^{t, (i)}, \theta^t, y^t)$ as a Gaussian distribution with mean vector 
$B^t x^{t, (i)} + C^t  y^t  + d^t$ and covariance matrix $ S^t$, 
\begin{align}
q (\tilde x^{t, (i)} | x^{t, (i)}, \theta^t, y^t)
=\mathcal N \left  (\tilde x^{t, (i)}; B^t x^{t, (i)} + C^t y^t  + d^t, S^t  \right ),
\label{eq:17}
\end{align}
where $B^t  \in \mathbb R^{n \times n}$, $C^t \in \mathbb R^{n \times m}$, $d^t \in \mathbb R^n$ 
and $S^t \in \mathbb R^{n \times n}$ are quantities that we need to decide so that Eq. \eqref{eq:11} is fulfilled. 
The $B^t$, $C^t$, $d^t$ and $S^t$ can all be functions of $\theta^t$ and $y^t$. 
From Eq. \eqref{eq:17}, it follows that the posterior sample $ \tilde x^{t, (i)}$ can be obtained as a linear shift of $x^{t, (i)}$ 
plus a zero-mean Gaussian noise term $\tilde \epsilon^{t, (i)} \sim \mathcal N(\tilde \epsilon^t; 0, S^t)$, 
\begin{equation}
\tilde x^{t, (i)} = B^t x^{t, (i)} + C^t y^t  + d^t + \tilde  \epsilon^{t, (i)}. 
\label{eq:18} 
\end{equation}
Using that $x^{t, (i)}$ in a similar fashion can be obtained as   $x^{t, (i)} = \mu^t + \omega^{t, (i)}$, where 
$\omega^{t, (i)} \sim \mathcal N(\omega^t; 0, Q^t )$, we can rewrite Eq.  \eqref{eq:18} as  
\[
\tilde x^{t, (i)} = B^t \mu^t + C^t y^t  + d^t+ B^t {\omega}^{t, (i)} +  \tilde {\epsilon}^{t, (i)}.
\]
Given $(\theta^t, y^t)$, the stochastic components on the right hand side of this equation are  ${\omega}^{t, (i)}$ and $ \tilde {\epsilon}^{t, (i)}$ which are independent and Gaussian. 
Thereby, since $\tilde x^{t, (i)}$ is  a linear combination of   ${\omega}^{t, (i)}$ and $ \tilde {\epsilon}^{t, (i)}$,  
we find that $\tilde x^{t, (i)} $ given $(\theta^t,y^t)$ is distributed according to a Gaussian distribution 
$\mathcal N(\tilde x^{t, (i)}; \tilde \mu^t, \tilde Q^t)$ with mean vector $\tilde \mu^t$ and covariance matrix $\tilde Q^t$ respectively given as 
\begin{equation}
\tilde \mu^t = B^t  \mu^t + C^t y^t + d^t
\label{eq:19}
\end{equation}
and
\begin{equation}
\tilde Q^t = B^t Q^t (B^t) ^\top + S^t. 
\label{eq:20}
\end{equation}

The requirement in Eq. \eqref{eq:10} now states that the mean vector $\tilde \mu^t $ in Eq. \eqref{eq:19} must be equal to $\mu^{*t} $ in Eq. \eqref{eq:4..}
and the covariance matrix $\tilde Q^t$ in Eq. \eqref{eq:20} must be equal to $Q^{*t}$ in Eq. \eqref{eq:5..}. That is, we must have
\begin{equation}
B^t \mu^t + C^ty^t + d^t = \mu^t + K^t(y^t-H^t\mu^t)
\label{eq:21}
\end{equation}
 and
 \begin{equation}
 B^t Q^t (B^t)^\top + S^t =  (I_n - K^tH^t)Q^t .
 \label{eq:22}
 \end{equation}
Solving Eq. \eqref{eq:21} with respect to $  C^ty^t + d^t$ and inserting the result into Eq.  \eqref{eq:18}, we obtain
 \begin{equation}
 \tilde x^{t, (i)} =  B^t(x^{t, (i)}  - \mu^t) + \mu^t + K^t(y^t-H^t\mu^t) +   \tilde  \epsilon^{t, (i)}. 
 \label{eq:23}
\end{equation}
Thereby, we see that in order to update $x^{t, (i)}$ we must  specify appropriate $B^t$ and $S^t$.  
To choose a procedure, one may either first choose $S^t$ and thereafter compute $B^t$ consistent with Eq. \eqref{eq:22}, 
or one may first choose $B^t$ and then compute $S^t$ consistent with Eq. \eqref{eq:22}.  
Below, we list some solutions that are particularly interesting.

\begin{example}
By choosing all elements of $B^t$ equal to zero, we obtain $\tilde x^{t, (i)}$ independent of $x^{t, (i)}$, 
\[
\tilde x^{t, (i)} = \mu^t + K^t (y^t-H^t\mu^t) +\tilde \epsilon^{t, (i)}.
\]
We then have $S^t= (I_n-K^tH^t)Q^t$, and  $q(\tilde x^{t, (i)} | x^{t, (i)}, \theta^t, y^t)$ is simply equal to the assumed posterior model $ f_{x^t |   \theta^t, y^t} (x^t |  \theta^t, y^t)$, i.e. the Gaussian distribution with mean and covariance given by Eqs.  \eqref{eq:4..} and \eqref{eq:5..}, respectively.
\end{example}

\begin{example}
By choosing all elements of $S^t$ equal to zero, the update of $x^{t, (i)}$  becomes deterministic and equivalent  to a square root EnKF. Specifically,  Eq. \eqref{eq:23} becomes equal to Eq. \eqref{eq:squareEnKF},  and Eq.  \eqref{eq:22} becomes equal to Eq.  \eqref{eq:7}. 
The distribution $ q(\tilde x^{t, (i)} | x^{t, (i)}, \theta^t, y^t)$ is then a degenerate Gaussian distribution, or a delta function. 
\end{example}

\begin{example}
\label{example3}
By choosing
\begin{equation}
\label{eq:24}
B^t = I_n-K^tH^t
\end{equation}
and
\begin{equation}
\label{eq:25}
S^t = (I_n-K^tH^t)Q^t(K^tH^t)^\top
\end{equation}
the update in Eq. \eqref{eq:23} becomes equivalent to the stochastic EnKF update in Eq. \eqref{eq:stochEnKF}.  
This result is proved in Appendix \ref{sec:A2}.
\end{example}

\subsection{The optimal solution}
\label{sec:4.3}
The optimality criterion we consider for this situation is to minimise the expected value of the Mahalanobis distance  $g(x^{t,(i)}, \tilde x^{t,(i)})$  in 
Eq. \eqref{eq:12} for a general positive definite matrix $\Sigma$. The minimisation is to be solved with respect 
to $B^t$ and $S^t$ under the restriction in Eq.  \eqref{eq:22} and, since $S^t$ is a covariance matrix, the additional restriction 
that $S^t$ is positive semidefinite.

To compute the optimal solution with respect to these criteria, we start out using that  $\Sigma^{-1}$ can be factorised as $\Sigma^{-1} = A^\top A$, $A \in \mathbb R^{n \times n}$. 
Hence,  the function to be minimised, with respect to $B^t$ and $S^t$, is
\begin{equation}
\text{E} \! \left [  g(x^{t,(i)}, \tilde x^{t,(i)}) \right ] = \text{E}\!\left [  \left (A(\tilde x^{t,(i)} - x^{t,(i)}) \right)^\top  \left  ( A(\tilde x^{t,(i)} - x^{t,(i)}) \right ) \right ], 
\label{eq:27}
\end{equation}
where the expectation is  taken over the  joint distribution $ f_{x^t|\theta^t}(x^t |\theta^t) q(\tilde x^{t,(i)} | x^t, \theta^t, y^t)$. 
Using Eq. \eqref{eq:23},  we can write $A(\tilde x^{t,(i)} - x^{t,(i)})$ as
\begin{align}
A (\tilde x^{t,(i)} - x^{t,(i)}) = A \left ( ( B^t - I_n)   (x^{t,(i)}  - \mu^t)  + K^t(y^t - H^t\mu^t) + \tilde{\epsilon}^{t,(i)} \right ). 
\label{eq:28}
\end{align}
Since  $\theta^t$ and $y^t$ are treated as  constants, the only stochastic components on the right hand side of Eq. \eqref{eq:28} 
are $x^t$ and $\tilde{\epsilon}^{t,(i)}$ which are independent and Gaussian. Thereby,  
$A (\tilde x^{t,(i)} - x^{t,(i)})$ is Gaussian since it is a linear
combination of independent Gaussian variables. Moreover, from Eq. 
(\ref{eq:28}) we see that 
\begin{equation*}
\mbox{E}\!\left[A(\tilde x^{t,(i)}-x^{t,(i)})\right] =
  AK^t(y^t-H^t\mu^t)
\end{equation*}
and
\begin{align*}
 \mbox{Cov}\!\left[A(\tilde x^{t,(i)}-x^{t,(i)})\right] 
=  A(B^t-I_n)Q^t(B^t-I_n)^\top A^\top + AS^tA^\top.
\end{align*}
Using that for any stochastic vector $w$ we have 
$\text{E}\!\left [ w^\top w  \right] = \text{tr}\!\left[\text{Cov} (w)\right]  + \mbox{E}[w]^\top \mbox{E}[w]$, 
we can write Eq. \eqref{eq:27} as 
\begin{align}
 & \mbox{E}\!\left[ \left(A(\tilde x^{t,(i)}-x^{t,(i)})\right)^\top
   \left(A(\tilde x^{t,(i)}-x^{t,(i)})\right)\right]  
  =\mbox{tr}\!\left(A(B^t-I_n)Q^t(B^t-I_n)^\top A^\top \right)+\mbox{tr}\!\left(AS^tA^\top\right) \label{eq:29} \\
& ~~~~~~~~~~~~~~~~~~~~~~~~~~~~~~~~~~~~~~~~~~~~~~~~~~~~~~~~~~~~
+\left(AK^t(y^t-H^t\mu^t)\right)^\top\left(AK^t(y^t-H^t\mu^t)\right).\nonumber
\end{align}
We see that the last term in this equation is constant as a function
of $B^t$ and $S^t$. Thereby,  to minimise Eq. \eqref{eq:27} with respect to $B^t$ and $S^t$ we only need to  minimise the sum of the two traces in Eq. \eqref{eq:29}. 
According to the restriction in Eq. \eqref{eq:22} we must have 
\begin{equation}
S^t = (I_n-K^tH^t)Q^t - B^t Q^t (B^t)^\top.
\label{eq:30}
\end{equation}
Using Eq. \eqref{eq:30},  we can write the sum of the two traces in Eq.  \eqref{eq:29} as a function of $B^t$ only,
\begin{eqnarray*}
 \text{tr} 
  \left \{
  A (B^t-I_n)
  Q^t (B^t-I_n)^\top A^\top
  \right \}
  +
   \text{tr}
  \left \{
  A S^t A^\top
  \right \}
=
\text{tr}
\left \{
-2AB^t Q^tA^\top +  2AQ^tA^\top - AK^t H^tQ^tA^\top
\right \} . 
\end{eqnarray*}
Here,  only the first term is a function of $B^t$. Hence,  minimising
Eq. \eqref{eq:27} with respect to $B^t$ is
equivalent to maximising 
\begin{eqnarray}
c(B^t) = \text{tr}\!\left \{   AB^t Q^t A^\top   \right \} 
\label{eq:31}
\end{eqnarray}
with respect to $B^t$ under the restriction that the matrix $S^t$ in Eq. \eqref{eq:30} is positive semidefinite.

To solve the optimisation problem stated above, we  first rephrase it to a standardised form. 
To do so, we start with singular value decompositions of the two covariance matrices $Q^t$ and  $(I_n-K^tH^t)Q^t$,
\begin{align}
&Q^t = VDV^\top, 
\label{eq:32} \\
&(I_n-K^tH^t)Q^t =  U \Lambda U^\top,
\label{eq:33}
\end{align}
where $U, V  \in \mathbb R^{n\times n}$ are orthogonal matrices, 
i.e. $UU^\top  = U^\top U=I$ and $VV^\top=V^\top V=I_n$,  and $ D, \Lambda  \in \mathbb
R^{n\times n}$ are diagonal matrices.
Inserting Eqs. \eqref{eq:32} and \eqref{eq:33} into Eq. 
\eqref{eq:30} and defining
\[
\tilde S^t = \Lambda^{-\frac{1}{2}}U^\top S^tU\Lambda^{-\frac{1}{2}}
\] and
\[
\tilde B^t = \left(\Lambda^{-\frac{1}{2}}U^\top B^tVD^{\frac{1}{2}}\right)^\top 
\]
we get that Eq. \eqref{eq:30} is equivalent to
\begin{equation}
\tilde S^t = I_n - (\tilde B^t)^\top \tilde B^t
\label{eq:34}
\end{equation}
and the objective function $c(B^t)$ in Eq.  \eqref{eq:31} can be
re\-phrased in terms of $\tilde B^t$ as 
\begin{align}\nonumber
  \tilde c(\tilde B^t) &= \mbox{tr}\!\left\{
  AU\Lambda^{\frac{1}{2}}(\tilde B^t)^\top D^{\frac{1}{2}} V^\top A^\top \right\}\\ \nonumber
  &= \mbox{tr}\!\left\{ \tilde
    B\Lambda^{\frac{1}{2}}U^\top A^\top AQ^tVD^{-\frac{1}{2}}\right\}\\
  &= \mbox{tr}\!\left\{ \tilde B^tZ^t\right\}, \label{eq:35}
\end{align}
where
\begin{equation}\label{eq:36}
  Z^t=\Lambda^{\frac{1}{2}}U^\top A^\top AQ^tVD^{-\frac{1}{2}}.
\end{equation}
Recognising that the matrix $\tilde S^t$ is positive semidefinite if and only
if $S^t$ is positive semidefinite, the rephrased optimisation problem is
thereby to maximise $\tilde c(\tilde B^t)$ in Eq. \eqref{eq:35} with respect to $\tilde B^t$  
under the constraint that $\tilde S^t$ in Eq. \eqref{eq:34} is positive semidefinite. 
To solve this standardised optimisation problem we can apply the following theorem for which a proof is given in Appendix \ref{sec:A1}.
\begin{theorem}
For a square matrix $Z \in \mathbb{R}^{n\times n}$ of full rank and with singular value decomposition $Z = PG F^\top$ the maximum value for 
$\text{tr} (\tilde B Z)$, $\tilde B \in \mathbb{R}^{n\times n}$ under the restriction
that $\tilde S = I_n-\tilde B^\top \tilde B$ is positive semidefinite occurs only for 
\[
\tilde B = F P^\top. 
\]
\label{thm}
\end{theorem}

To apply Theorem \ref{thm} we first need to argue why the matrix $Z^t$ in Eq. \eqref{eq:36} has full rank. 
Since $Q^t$ and $(I_n-K^tH^t)Q^t$ are positive definite matrices, $D$ and $\Lambda$ are invertible. 
Thereby also $D^{\frac{1}{2}}$ and $\Lambda^{\frac{1}{2}}$ are invertible. 
$V$ and $U$ are both orthogonal and thereby invertible. 
Finally, as we have required $\Sigma$ to be positive definite, $\Sigma$ is invertible, and when $\Sigma$ is invertible, $A$ is also invertible. 
Thereby, $Z^t$ is given as a product of invertible matrices and is therefore itself invertible and has full rank. 
 
According to Theorem \ref{thm} the solution to our optimisation problem in standardised form is $\tilde B^t = FP^\top $. 
We thereby get that 
\[
\tilde S^t = I_n - (FP^\top)^\top FP^\top = I_n - P F^\top F P^\top = 0, 
\] 
i.e. all elements in $\tilde S^t$, and hence all elements in $S^t$, are zero. 
The solution to our optimisation problem thereby corresponds to a square root EnKF. 
The corresponding optimal value for $B^t$ is 
\[
B^t= U \Lambda^{\frac{1}{2}} P F^\top D^{-\frac{1}{2}} V^\top. 
\]

\subsection{Parameter simulation}
\label{sec:4.4}
According to step 2a) in Algorithm \ref{alg:1}, we need to simulate  $\theta^{t, (i)} | x^{t, -(i)}, y^t \sim f_{\theta^t | x^{t, -(i)}, y^t}(\theta^t | x^{t, -(i)}, y^t)$ prior to constructing $q(\tilde x^{t, (i)} | x^{t, (i)}, \theta^t, y^t)$. 
For this, we can construct a Gibbs sampler as explained in  Section \ref{sec:3.4}. 
For the linear-Gaussian model we now consider, we have $\theta^t = (\mu^t, Q^t)$. 
To construct the Gibbs sampler we need to derive the full conditional distributions $f_{x^t|\theta^t,  y^t}(x^t|\theta^t, y^t)$ and $f_{\theta^t | x^t, x^{t, -(i)}}(\theta^t | x^t, x^{t, -(i)})$.
From previous sections, we know that  $f_{x^t |\theta^t, y^t}(x^t |  \theta^t, y^t)$ is a Gaussian distribution, $\mathcal N(x^t; \mu^{*t}, Q^{*t})$, 
with parameters $\mu^{*t}$ and $Q^{*t}$ given by Eqs. \eqref{eq:4..} and \eqref{eq:5..}, respectively. 
To simulate from  $f_{\theta^t | x^t, x^{t, -(i)}}(\theta^t | x^t, x^{t, -(i)})$, we first factorise it as 
$$
f_{\theta^t | x^t, x^{t, -(i)}}(\theta^t | x^t, x^{t, -(i)}) = f_{Q^t | x^t, x^{t, -(i)}}(Q^t | x^t, x^{t, -(i)}) f_{\mu^t | Q^t, x^t, x^{t,-(i)}}(\mu^t | Q^t, x^t, x^{t,-(i)}). 
$$
Since conjugate priors are chosen for $\mu^t$ and $Q^t$, it can be shown that  
$f_{Q^t | x^t, x^{t, -(i)}}(Q^t | x^t, x^{t, -(i)}) $ is  an inverse Wishart distribution, 
\[
f_{Q^t | x^t, x^{t, -(i)}}(Q^t | x^t, x^{t, -(i)}) = \mathcal W^{-1} (Q^t;  \tilde V, \tilde \nu), 
\]
where
\[
\tilde \nu = \nu + M
\]
and 
\[
\tilde V = V +  C^{t, (i)} + \frac{\kappa M}{\kappa + M} \left ( \bar x^{t, (i)} -\mu_0 \right ) \left (  \bar x^{t, (i)} - \mu_0 \right )^\top,
\]
where 
\[
\bar x^{t, (i)} = \frac{1}{M} \left ( x^t +  \sum_{j\neq i} x^{t, (j)} \right )
\]
and 
\begin{align*}
C^{t, (i)} = \left ( x^t - \bar x^{t, (i)} \right ) \left ( x^t - \bar x^{t, (i)}
                 \right ) ^\top
  + \sum_{j \neq i} \left ( x^{t, (j)} - \bar x^{t, (i)} \right ) \left ( x^{t, (j)} - \bar x^{t, (i)} \right ) ^\top,
\end{align*}
and $f_{\mu^t | Q^t, x^t, x^{t,-(i)}}(\mu^t | Q^t, x^t, x^{t,-(i)})$ is a Gaussian distribution, 
$$
f_{\mu^t | Q^t, x^t, x^{t,-(i)}}(\mu^t | Q^t, x^t, x^{t,-(i)}) = \mathcal N (\mu^t; \tilde \mu_0, \tilde \kappa^{-1} Q^t),
$$
where
\[
\tilde \mu_0 = \frac{\kappa \mu_0 + M \bar x^{t, (i)}}{\kappa + M}
\]
and 
\[
\tilde \kappa = \kappa + M. 
\]

\section{Application 2: First-order Markov chain assumed model}
\label{sec:5}
In this section, we describe how the general updating procedure described in Section \ref{sec:3} can be applied when the elements of the state vector $x^t$ are categorical variables, $x_j^t \in \{0, 1, \dots, K-1\}$,  and $x^t$ is restricted to have a one-dimensional spatial arrangement. 
We then propose to let   $f_{x^t|\theta^t}(x^t|\theta^t)$ and $f_{y^t|x^t}(y^t|x^t)$ constitute  a hidden Markov model (HMM). 
 The following material can be seen as a generalised and fully Bayesian version 
of the updating method for binary state vectors proposed in \cite{LoeTjelmeland2020}. 

\subsection{Specification of the assumed model}
\label{sec:5.1}
Suppose $x^t = (x_1^t, \dots, x_n^t)$ is a vector of $n$ categorical variables, $x_j^t \in \{0,1, \dots, K-1\}$, and 
suppose $x^t$ has a one-dimensional spatial arrangement (i.e., the vector is spatially arranged along a line).  A natural choice of model for $f_{x^t|\theta^t}(x^t | \theta^t)$ is then a first-order Markov chain,
\begin{equation}
f_{x^t|\theta^t}(x^t | \theta^t)  = f(x_1^t | \theta^t) \prod_{j=2}^n f( x_j^t | x_{j-1}^t, \theta^t).
\label{eq: Markov chain}
\end{equation}
Moreover, suppose $y^t = (y_1^t, \dots, y_n^t)$ is a vector of $n$ variables, $y_j^t \in \mathbb R$, so that 
we have  one observation $y_j^t$ for each component $x_j^t$ of $x^t$, and assume that the $y_j^t$'s are 
conditionally independent given $x^t$,
\[
f_{y^t|x^t}(y^t|x^t) = \prod_{j=1}^n f_{y^t_j|x_j^t}(y_j^t | x_j^t). 
\]
Given $\theta^t$, the pair $f_{x^t|\theta^t}(x^t|\theta^t)$ and $f_{y^t|x^t}(y^t|x^t)$ constitute a HMM. 
The corresponding posterior model $f_{x^t|\theta^t, y^t}(x^t|\theta^t, y^t)$ is then also a first-order Markov chain 
whose initial and  transition probabilities can be computed with the  the forward-backward algorithm for HMMs (e.g., \citealp{chapter1}).

The  parameter $\theta^t$ may in this context represent the initial and transition probabilities  of the assumed first-order Markov chain $f_{x^t|\theta^t}(x^t | \theta^t)$.  
In the following, we let
\[
\theta^t = \left ( \{ \theta^t_1(i) \}_{i=0}^{K-1}, \{ \theta_2^{t,k}(i) \}_{i,k=0}^{K-1}, \dots, \{\theta_{n}^{t,k}(i)\}_{i,k=0}^{K-1}  \right  ),
\]
where $\theta_1(i)^t, \theta_j^{t,k}(i) \in (0,1)$, $\sum_{i=0}^{K-1} \theta_j^{t,k}(i) = 1$,
and
\[
 f(x_1^t = i |\theta^t) = \theta_1^t(i)
\]
and
\[
f(x_j^t = i | x_{j-1}^t = k , \theta^t) = \theta_j^{t,k}(i), 
\]
for $ i, k = 0, \dots, K-1$ and $j = 2, \dots, n.$ For convenience, we also define
\[
\theta_1^t = ( \theta_1^t(0), \theta_1^t(1),  \dots, \theta_1^t(K-1) )
\]
and
\[
\theta_j^{t,k} = (\theta_j^{t,k}(0),\theta_j^{t,k}(1),  \dots, \theta_j^{t,k}(K-1)). 
\]
As discussed in Section \ref{sec:3.4},  the prior  $f_{\theta^t}(\theta^t)$ should be chosen as conjugate for $f_{x^t|\theta^t}(x^t|\theta^t)$.
To obtain this,  we first assume that all the vectors $\theta_1^t$, $\theta_2^{t,0}, \dots, $ $\theta_2^{t, K-1},$ 
$\theta_3^{t,0}, \dots,$ $\theta_3^{t,K-1}, \dots, \theta_n^{t,0}, \dots, \theta_n^{t,K-1}$ are a priori independent, so that
\[
f_{\theta^t}(\theta^t) = f_{\theta_1^t}(\theta_1^t) \prod_{j,k} f_{\theta_j^{t,k}}(\theta_j^{t,k}), 
\]
and then choose $f_{\theta_1^t} (\theta_1^t) $ as  a Dirichlet distribution with parameters $\alpha_1^t(0), \dots, \alpha_{1}^t(K-1)$,
\[
f_{\theta_1^t} (\theta_1^t) \propto \prod_{i=0}^{K-1} \theta_1^t(i),
\]
 and choose each $f_{\theta_j^{t,k}}(\theta_j^{t,k})$ as  a Dirichlet distribution with parameters $\alpha_j^{t,k}(0), \dots, \alpha_j^{t,k}(K-1)$,
\[
f_{\theta_j^{t,k}}(\theta_j^{t,k}) \propto \prod_{i=0}^{K-1} \theta_{j}^{t,k}(i) ^{\alpha_j^{t,k}(i)}.
\]
The hyperparameters $\alpha_1^t(i)$ and $\alpha_j^{t,k}(i)$, $i, k=0, \dots, K-1$,  $j = 2, \dots, n$, are all assumed to be  known.

\subsection{Class of updating distributions}
\label{sec:5.2}

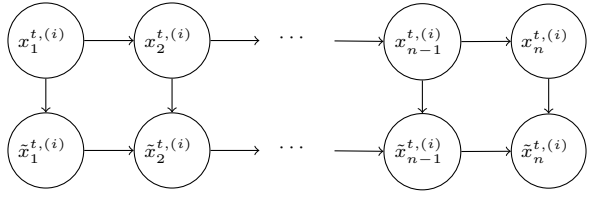
\begin{figure*}
	\centering
	\begin{tikzpicture}[
	align = flush center,
	font = \footnotesize]
	\matrix [
	matrix of nodes,
	column sep = 2em,
	row sep = 1.3em, 
	nodes = {solid},
	] (m)
	{  
		  |[data]|$x_1^{t, (i)}$ & |[data]|$x_2^{t, (i)}$ &  |[empty]| \raisebox{0.07cm}{$\cdots$} &  |[data]|$x_{n-1}^{t, (i)}$ &  |[data]|$x_n^{t, (i)}$  \\
		 |[data]|$\tilde x_1^{t, (i)}$ &  |[data]|$\tilde x_2^{t, (i)}$ & |[empty]| \raisebox{0.07cm}{$\cdots$} &  |[data]|$\tilde x_{n-1}^{t, (i)}$  &  |[data]|$\tilde x_n^{t, (i)}$  \\
		\\[.7em]
	}; 
	\draw[->](m-1-1) to (m-1-2);
	\draw[->](m-1-2) to (m-1-3);
	\draw[->](m-1-3) to (m-1-4);
	\draw[dotted](m-1-4);
	\draw[->](m-1-4) to (m-1-5);
	
	\draw[->](m-1-1) to (m-2-1);
	\draw[->](m-1-2) to (m-2-2);
	\draw[->](m-1-4) to (m-2-4);
	\draw[->](m-1-5) to (m-2-5);
	
	\draw[->](m-2-1) to (m-2-2);
	\draw[->](m-2-2) to (m-2-3);
	\draw[->](m-2-3) to (m-2-4);
	\draw[dotted](m-1-4);
	\draw[->](m-2-4) to (m-2-5);
	
	\end{tikzpicture}
	\caption{Graphical illustration of enforced dependencies between the variables in a prior sample $x^{t, (i)}$ and corresponding posterior sample $\tilde x^{t, (i)}$, given $\theta^t$ and $y^t$, in the first-order Markov chain application of the proposed updating approach.}
	\label{fig:q}
\end{figure*}

Because of the discrete context of the current situation, the criterion in Eq. \eqref{eq:11} can be written as a sum, 
\begin{align}
  f_{x^t|y^t,\theta^t}(\tilde x^{t, (i)} | y^t, \theta^t)
  = \sum_{x^{t, (i)} \in \Omega_x} f_{x^t|\theta^t} (x^{t, (i)}|\theta^t) q(\tilde x^{t, (i)} | x^{t, (i)}, \theta^t, y^t). 
\label{eq: restriction discrete}
\end{align}
Brute force,  the updating distribution $q(\tilde x^{t, (i)} | x^{t, (i)}, \theta^t, y^t)$ now represents a transition matrix, and  there are $K^n ( K^n-1)$ transition probabilities that need to be specified.
Even when $n$ is only moderately large this becomes too computationally demanding.  
To simplify the situation,  we therefore enforce a certain dependency structure for  
 $q(\tilde x^{t, (i)} | x^{t, (i)}, \theta^t, y^t)$ 
 as illustrated in Figure \ref{fig:q}. We can then factorise 
    $q(\tilde x^{t, (i)} | x^{t, (i)}, \theta^t, y^t)$  as
\begin{align}
q(\tilde x^{t, (i)} | x^{t, (i)}, \theta^t, y^t) &= q(\tilde x^{t, (i)}_1 |
                                         x_1^{t, (i)}, \theta^t, y^t)
   \prod_{j=2}^n q(\tilde x^{t, (i)}_j | \tilde x^{t, (i)}_{j-1}, x^{t, (i)}_j,  \theta^t, y^t). 
\label{eq:q fact}
\end{align}
The number of quantities required to specify $q(\tilde x^{t, (i)} | x^{t, (i)}, \theta^t, y^t)$ thereby reduces to $K(K-1) + (n-1)K^2(K-1)$, or more specifically  $K(K-1)$ quantities for $q(\tilde x_1^{t, (i)} | x_1^{t, (i)}, \theta^t, y^t)$ and  $K^2(K-1)$ quantities for each factor $q(\tilde x_j^{t, (i)} | \tilde x_{j-1}^{t, (i)}, x_j^{t, (i)}, \theta^t, y^t)$, $j = 2, \dots, n.$ As this is a linear, rather than an  exponential,  function of $n$, $n$ can be large without causing trouble. 

According to the requirement in Eq. \eqref{eq: restriction discrete},  $q(\tilde x^{t,(i)} | x^{t,(i)}, \theta^t, y^t)$ must be constructed such that marginalising out  $x^{t,(i)}$ from the joint distribution $f_{x^t|\theta^t} (x^{t,(i)}|\theta^t) q(\tilde x^{t, (i)} | x^{t,(i)},\theta^t, y^t)$ returns 
the posterior Mar\-kov chain model $f_{x^t|\theta^t, y^t}(\tilde x^{t,(i)} |\theta^t, y^t)$. However, the problem of constructing such a 
$q(\tilde x^{t,(i)} | x^{t,(i)}, \theta^t, y^t)$, different from $f_{x^t|\theta^t, y^t}(x^t|\theta^t, y^t)$ itself, is generally  too intricate to solve. 
Therefore, we need to settle with an approximate approach. As in \cite{LoeTjelmeland2020}, we propose to replace the requirement of retaining
the whole Markov chain model $f_{x^t| \theta^t, y^t}(x^t |  \theta^t, y^t)$ 
with the requirement that only the bivariate probabilities 
$f_{x_j^t, x^t_{j+1} | \theta^t, y^t}({x^t_j, x^t_{j+1} | \theta^t, y^t})$ are retained, i.e. 
\begin{align}
f_{\tilde x^{t,(i)}_j, \tilde x^{t,(i)}_{j+1} | \theta^t, y^t} (x^t_j,  x^t_{j+1} | \theta^t, y^t)
= f_{x_j^t, x^t_{j+1} | \theta^t, y^t} (x^t_j, x^t_{j+1} | \theta^t, y^t), \quad j=1,\ldots,n-1. 
\label{eq: marginal requirement}
\end{align}
This means that, under the assumption that the assumed model is correct, the distribution of the updated sample 
$\tilde x^{t,(i)}$ given  $(\theta^t, y^t)$ is not equal to the first-order Markov chain  $f_{x^t| \theta^t, y^t}(x^t | \theta^t, y^t)$, but that each pair  $(\tilde x_j^{t,(i)}, \tilde x^{t,(i)}_{j+1}), j = 1, \dots, n-1,$ is marginally distributed according to the bivariate distribution
$f_{x_j^t, x_{j+1}^t | \theta^t, y^t} (x_j^t, x_{j+1}^t | \theta^t, y^t)$ of the Markov chain. 

\subsection{The optimal solution}
\label{sec:5.3}

The optimality criterion we consider for this situation is to minimise the expected number of components of 
$x^{t, (i)}$ that are different from their corresponding components in $\tilde x^{t, (i)}$; that is,  we want to minimise the expected value of  the function 
$g(x^{t, (i)}, \tilde x^{t, (i)})$ in Eq.  \eqref{eq:14}. 
Minimising $\text{E} \left  [ g(x^{t,(i)}, \tilde x^{t,(i)}) \right ]$  is then equivalent to maximising
\begin{equation}
\text{E} \left [ \sum_{j=1}^n 1(x_j^{t,(i)} = \tilde x_j^{t,(i)}) \right ]
\label{eq: objective}
\end{equation}
where the expectation is taken over $f_{x^t|\theta^t}(x^t|\theta^t) q(\tilde x^{t,(i)} | x^t, \theta^t, y^t)$. 
We are thereby faced with a constrained optimisation problem where we want to maximise, with respect to $q(\tilde x^{t,(i)}| x^{t,(i)}, \theta^t, y^t)$, the function in Eq. \eqref{eq: objective}
under the condition in Eq. \eqref{eq: marginal requirement} and under the condition that $q(\tilde x^{t,(i)} | x^{t,(i)}, \theta^t, y^t)$ can be factorised as in Eq. \eqref{eq:q fact}. 

\cite{LoeTjelmeland2020} propose a dynamic programming algorithm for solving the optimisation problem stated above when $x_j^t$ is binary, $x_j^t \in \{0,1\}$. 
The proposed algorithm is based on that the maximum value of Eq.  \eqref{eq: objective} can be computed recursively since
\begin{align}\nonumber
\max_{q_{k:n}^t} \text{E}\!\left [ \sum_{j=k}^n 1(x_j^{t,(i)} = \tilde x_j^{t,(i)}) \right ] 
= \max_{q_{k:n}^t} \text{E}\!\left [ 1(x_k^{t,(i)} = \tilde x_k^{t,(i)}) + \sum_{j=k+1}^n 1(x_j^{t,(i)} = \tilde x_j^{t,(i)}) \right ] 
\\
= \max_{q_k^t} \text{E}\!\left [ 
1(x_k^{t,(i)} = \tilde x_k^{t,(i)})  + \max_{q_{k+1:n}^t } \text{E}\!\left [ \sum_{j=k+1}^n 1(x_j^{t,(i)} = \tilde x_j^{t,(i)}) \right ]   
\right ]
\label{eq: backward}
\end{align}
where $q_k^t = q(\tilde x_k^{t,(i)} | \tilde x_{k-1}^{t,(i)}, x_k^{t,(i)}, \theta^t, y^t)$, $q_1^t = q(\tilde x_1^{t,(i)} | x_1^{t,(i)}, \theta^t, y^t)$, and $q_{k:n}^t = (q_k^t, \dots, q_n^t)$. 
The algorithm starts with a 'backward' recursion where,  for $k=n, n-1, \dots, 1$,  Eq.  \eqref{eq: backward} 
and the optimal value of $q_k^t$ are computed as functions of $q_{1:k-1}^t = (q_1^t, \dots, q_{k-1}^t)$. 
At the final step of the backward recursion  the whole expectation in Eq.  \eqref{eq: objective} is thereby computed, along with the optimal value for $q_1^t$. 
The algorithm then proceeds with a 'forward' recursion where, for $k=2,
\dots, n$, we recursively compute the optimal values for $q_2^t, \dots,
q_{n}^t$. Using linear programming, we are currently in the process of
developing an alternative algorithm for solving the optimisation problem 
when the number of possible values of $x_j^t$ is larger than two.

\subsection{Parameter simulation}
\label{sec:5.4}
To construct the Gibbs sampler described in Section \ref{sec:3.4} for simulating $\theta^{t, (i)} | x^{t, -(i)}, y^t $ 
we need to be able to simulate from the distributions 
$f_{x^t | \theta^t, y^t}(x^t | \theta^t, y^t)$ and  $f_{\theta^t | x^t, x^{t, -(i)}}(\theta^t | x^t, x^{t, -(i)})$. 
From Section \ref{sec:5.1} we know that 
$f_{x^t | \theta^t, y^t}(x^t | \theta^t, y^t)$ now 
is a first-order Markov chain with transition probabilities that are easy to compute with the forward-backward algorithm for HMMs. 
When it comes to $f_{\theta^t | x^t, x^{t, -(i)}}(\theta^t | x^t, x^{t, -(i)}) $, it can easily  be shown that $\theta_1^t | x^t, x^{t, -(i)}$ is Dirichlet distributed with parameters 
\[
\tilde {\alpha}_1^t(r) = \alpha^t_1(r) +  1(x_1^t = r) + \sum_{m \neq i} 1 \left (x_1^{t, (m)} = r \right ), 
\]
for $r = 0, \dots, K-1$. Similarly, it can be shown that  each $\theta_j^{t,k} | x^t, x^{t, -(i)}$ is Dirichlet distributed with parameters 
\begin{align*}
\tilde {\alpha}_{j}^{t,k}(r) = \alpha_j^{t,k}(r) + 1(x_{j-1} = k, x_{j} =
                 r)
  + \sum_{m \neq i} 1 \left (x^{t, (m)}_{j-1} = k, x^{t, (m)}_{j} = r \right )
\end{align*}
for $r = 0, \dots, K-1$. Moreover, all the parameters are independent a posteriori, 
\[
f_{\theta^t | x^t, x^{t, -(i)}}(\theta^t | x^t, x^{t, -(i)}) = f_{\theta_1^t  | x^t, x^{t, -(i)} }(\theta_1^t  | x^t, x^{t, -(i)} ) \prod_{j,k} f_{\theta_j^{t,k}  | x^t, x^{t, -(i)} }(\theta_j^{t,k}  | x^t, x^{t, -(i)} ). 
\]

\section{Simulation experiment with a linear-Gaussian  assumed model}
\label{sec:6}

In this section,  we present a simulation experiment for the situation described in Section \ref{sec:4}. We adopt an experimental setup previously 
used in \citet{art20}. In the following,  we first describe how we generate a
reference time series and simulate corresponding observations. Thereafter, 
we specify the precise assumed model we are using, and finally we present
and discuss simulation results.

\subsection{Experimental setup}
\label{sec:6.1}
To generate a reference time series $\{x^t\}_{t=1}^T$ that we consider as the true unobserved
state process we adopt the same setup as in \citet{art20}. At each time
$t$,  we assume that the state vector $x^t=(x^t_1,\ldots,x^t_n)$ consists of $n=100$  
continuous variables so that  $\Omega_x=\mathbb{R}^{100}$. The latent
process is defined from time $1$ to time
$T=11$. The values of the initial state vector, $x^1$, is generated
from a Gaussian distribution with zero mean, where the variance
of each component is $20$ and where the correlation between elements
$r$ and $s$ in $x^1$ is
\begin{equation}
c(r,s) = \exp\left\{-\frac{3|r-s|}{20}\right\}.
\end{equation}
\citet{art20} define two deterministic ways to generate $x^t,t=2,\ldots,T$
from $x^1$, one linear forward function and one non-linear. We adopt
the same linear forward function as used there, but not the same
non-linear function. The non-linear forward
function used in \citet{art20} induces a light-tailed bi-modal marginal
distribution for each component in the state vector at time $t=T$. We
construct instead a forward function which produces a heavy-tailed
one-mode marginal distribution for time $t>1$.

For $t=2,\ldots,T$, the linear forward function we use is defined by
\begin{equation}
x^t = \xi^{t-1} x^{t-1},
\end{equation}
where $\xi^{t-1}$ is an $n\times n$ matrix defined so that for $j=5t-4,\ldots,5t+5$,
element $j$ in $x^t$ is set equal to the average of elements $\max\{1,j-4\}$ to
$j+5$ in $x^{t-1}$, whereas the remaining elements in $x^t$ equal the corresponding
elements in $x^{t-1}$. The effect of this forward function is that the first part of
the vector $x^t$ is a smoothed version of the first part of $x^1$,
whereas the rest of $x^t$ equals the corresponding part of $x^1$.
When the time $t$ increases, the part that has been
smoothed also increases.

For the non-linear forward function, we simply transform the Gaussian
distributed elements in the state vector at time $t=1$ to be from a (scaled)
$t$-distribution at any later time $t>1$.
More specifically, element $j$ in $x^2$ is defined from the
corresponding element in $x^1$ by 
\begin{equation}\label{eq:forwardNonlinear1}
  x^2_j = \sqrt{20} F^{-1}_{\mathcal T}\!\left(\!\Phi\left(\frac{x^1_j}{\sqrt{20}}\right),100\right),
\end{equation}
where $F_{\mathcal T}(\cdot,\nu)$ and $\Phi(\cdot)$ are the cumulative
distribution functions for a $t$-distribution with $\nu$ degrees of
freedom and a standard normal distribution, respectively. Thus, the
marginal distribution of each element in $x^2$ is a $t$-distribution
with $100$ degrees of freedom. For later times $t>2$, each element $j$
in $x^{t}$ is defined from the corresponding element in $x^{t-1}$ by
\begin{equation}
x^{t}_j = \sqrt{20} F^{-1}_{\mathcal T}\left(
            F_{\mathcal T}\left(\frac{x^{t-1}_j}{\sqrt{20}},\nu_{t-1}\right),\nu_t\right),
\label{eq:forwardNonlinear2}
\end{equation}
where $\nu_t = 100/(2t-3)$.
Thus, the marginal distribution for each element gets heavier and
heavier tails when the time $t$ increases.

\begin{figure*}
  \begin{center}
    \begin{tabular}{@{}c@{}c@{}}
      \includegraphics[width=0.35\textwidth]{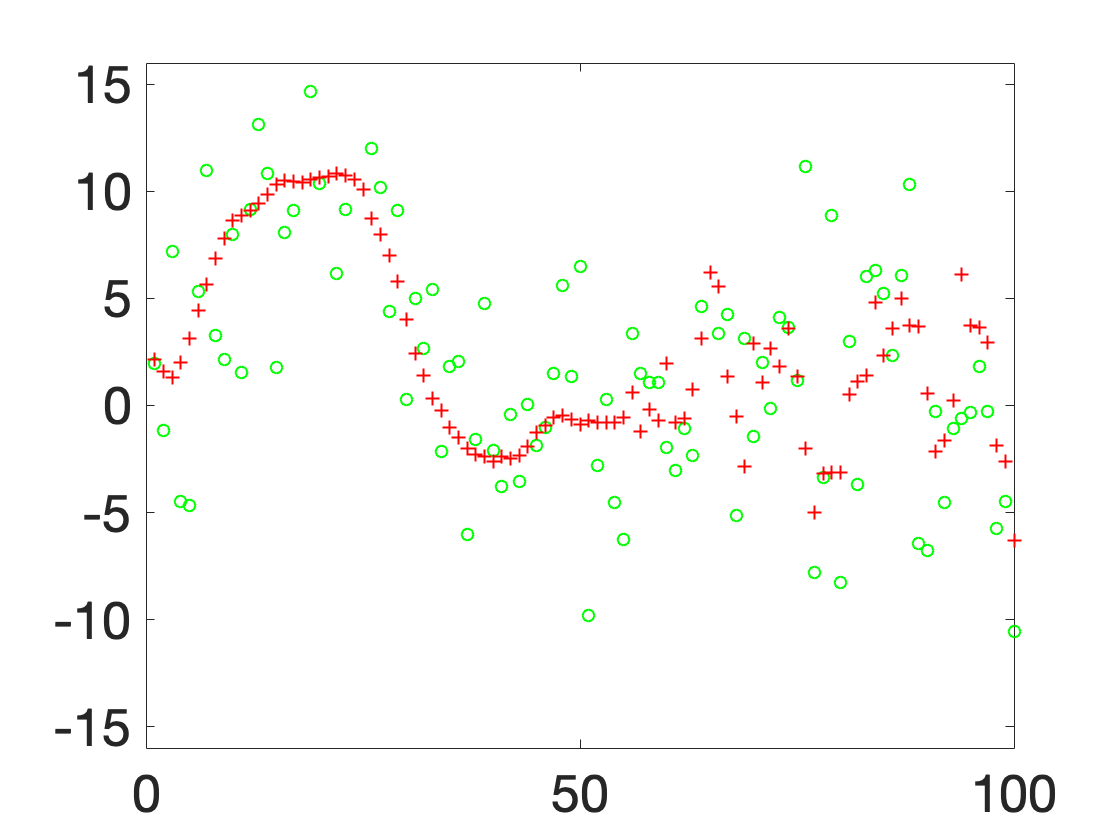}&
      \includegraphics[width=0.35\textwidth]{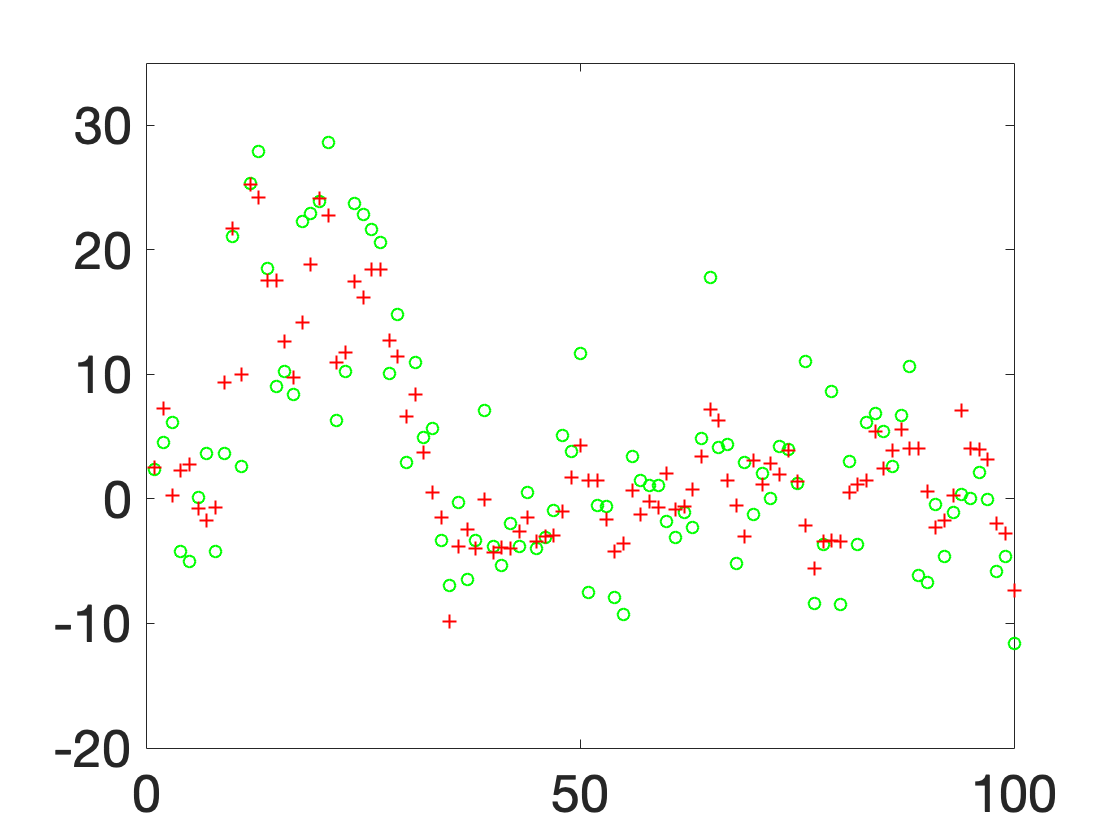}\\
      (a) & (b)\\[-0.3cm]
    \end{tabular}
  \end{center}
  \caption{\label{fig:refobs}The reference state vector (red crosses) at time $t=T$
    for the (a) linear and (b) non-linear forward model cases, and the
    simulated observations (green circles) at the same time. Note that a few of the
    observations are outside the range of the vertical axis.}
  
\end{figure*}

Having generated a reference time series $\{x^t\}_{t=1}^T$ as described above,
observations are simulated for each time $t=1,\ldots,T$. For each time
$t=1,\ldots,T$ an observation vector $y^t$ is simulated according to
\begin{equation}\label{eq:assumedLikelihood}
  y^t|x^t \sim {\cal N}\left(y^t; x^t,20I_n\right). 
\end{equation}
The reference state vectors for
the linear and the non-linear models at time $t=T$ and the corresponding simulated
observations at that time step are shown in Figure \ref{fig:refobs}.

\subsection{Details of the assumed model}
\label{sec:6.2}
The assumed model is as specified in Section
\ref{sec:4.1}.  The hyperprior
 in Eqs. (\ref{eq:15}) and (\ref{eq:16})  for
$\theta^t = (\mu^t, Q^t)$  is specified by four hyperparameters:
$\mu_0,\kappa,\nu$ and $V$. We choose values for these hyperparameters to
get a vague, but proper prior for $\theta^t$, and use the same values
for all time steps. We set all the elements of $\mu_0\in\mathbb{R}^n$
equal to zero, and set $\kappa=10$, $\nu = n+1.1$ and $V=(\nu - n - 1)
I_n$. Note that this in particular gives $\mbox{E}[Q^t]=I_n$ a priori.
For the likelihood $f_{y^t|x^t}(y^t|x^t)$ we use the same distribution as the one we
used to simulate the data, i.e. $f_{y^t|x^t}(y^t|x^t)$ is specified by Eq. 
(\ref{eq:assumedLikelihood}).

\subsection{Simulation results}
\label{sec:6.3}
When evaluating the performance of the proposed approach, the results
are compared with several other variants of EnKF. When updating one of
the ensemble members, there are two important steps. The first step is
how to generate or estimate $\mu^t$ and $Q^t$ based on the prediction
ensemble. The second step is how to use these $\mu^t$ and $Q^t$ values to
update the ensemble member in question. We consider tree variants of
the first step. The first is what we propose in this report, to sample $\mu^t$ and $Q^t$
from a posterior distribution given the new observation $y^t$ and
all ensemble members,  except the member which is to be
updated. For the function $g(x^{t, (i)},\tilde x^{t, (i)})$ we here use the
Eucledian distance, i.e. $\Sigma=I_n$.
The second is what \citet{art20} are advocating, to sample
$\mu^t$ and $Q^t$ from a posterior distribution given all the ensemble
members, including also the member that is going to be updated, but
not given the new observation $y^t$. The third is the standard procedure
in EnKF, to estimate $\mu^t$ and $Q^t$ based on all the ensemble members.
For how to update an ensemble member when values of $\mu^t$ and $Q^t$ are
given, we consider two variants. The first is the square-root filter
we found to be optimal in Section \ref{sec:4.3} and the
second is the standard stochastic EnKF update procedure specified in Eq. 
(\ref{eq:stochEnKF}). By combining each of the three variants of
how to generate $\mu^t$ and $Q^t$ with each of the two variants of how to
update the ensemble members, one can define six updating
procedures. We present results for all the six combinations.

\begin{figure*}
  \begin{center}
    \begin{tabular}{@{}c@{}c@{}}
      \includegraphics[width=0.35\textwidth]{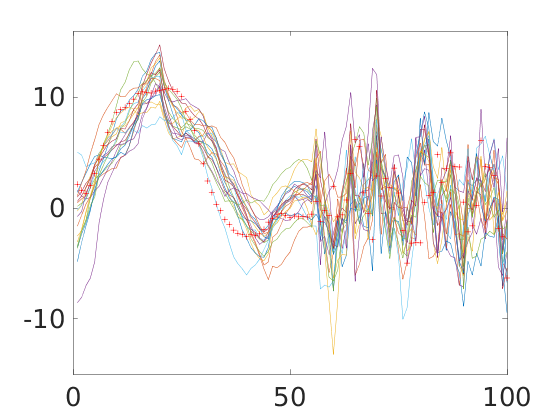}
      &
        \includegraphics[width=0.35\textwidth]{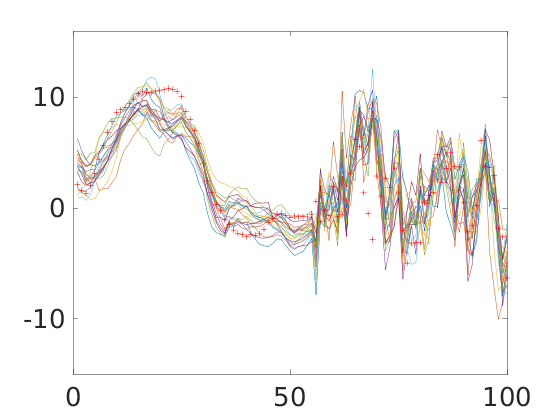}
      \\ \includegraphics[width=0.35\textwidth]{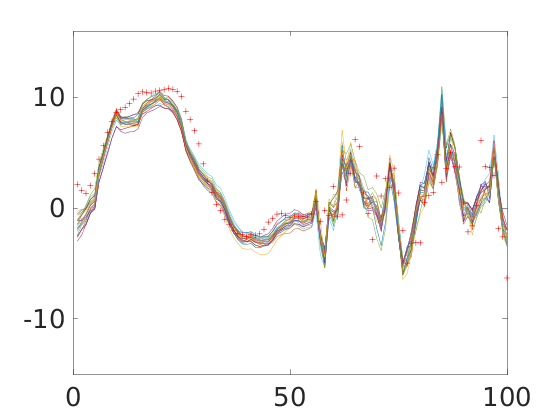}
      &
        \includegraphics[width=0.35\textwidth]{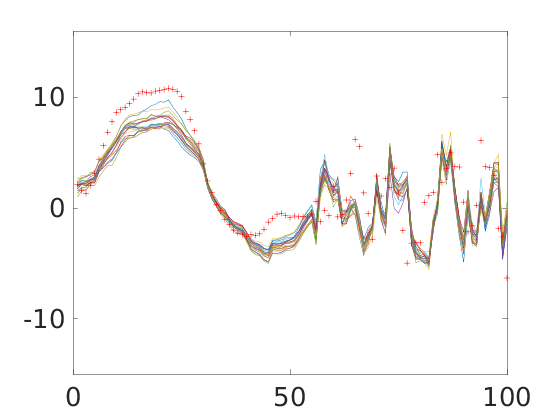}
      \\ \includegraphics[width=0.35\textwidth]{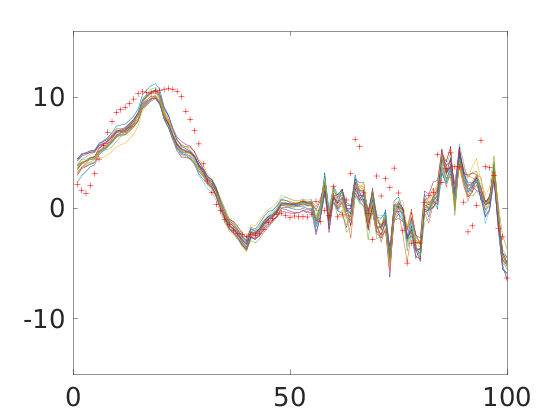}
      &
        \includegraphics[width=0.35\textwidth]{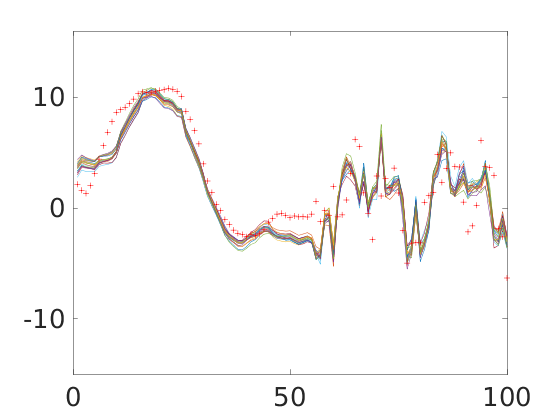}
      \\[-0.2cm]
    \end{tabular}
  \end{center}
  \caption{\label{fig:GaussianLinearState}Gaussian linear example:
    Prediction ensemble at time $T=11$ when using $M=19$ ensemble
    members. The upper, middle and lower rows are when using our proposed procedure
  for generating $\mu^t$ and $Q^t$, when using the procedure of
  \citet{art20} for the same, and when using empirical estimates,
  respectively. The left and right columns are when updating with our
  optimal square-root filter and when using the standard stochastic
  EnKF procedure, respectively. The ensemble members are shown with
  solid lines and the latent true state is shown with red crosses.}
\end{figure*}

Using the linear forward model described in Section
\ref{sec:6.1}, the prediction ensembles at time $T=11$
in one run of each of the six procedures considered,  with $M=19$
ensemble members, are shown in
Figure \ref{fig:GaussianLinearState}.
The ensemble members, drawn with solid lines in the figure, should thus
be considered as (approximate) samples from the distribution
$p_{x^{11}|y^{1:10}}(x^{11}|y^{1:10})$. For comparison, the latent true state vector at
time $T=11$ is also shown, with red crosses. The upper, middle and lower
lines show results when using our proposed procedure for generating $\mu^t$ and
$Q^t$, when using the procedure in \citet{art20} for the same, and when
using empirical estimates, respectively. The left and right columns
show results when using our optimal square-root filter to update the
ensemble members, and when using the standard stochastic EnKF update,
respectively.

The most striking difference between the six cases is the spread of
the ensemble members. In the four lower figures the spread is very
small, and as a result the latent true value is in most places
outside the spread of the ensemble members. For the standard
stochastic EnKF procedure, shown in the lower right figure, this
should come as no surprise as it is well known that this procedure tends to 
underestimate the uncertainty. 
What is more surprising is that the increase of the spread is so small
when instead using the procedure proposed in \citet{art20}, shown in
the middle right figure. The difference in the spread of the ensemble
members in each of the figures in the middle row and the corresponding
figure in the upper row is also striking, when remembering the very small
difference in the procedures used to generate the figures. The only
difference between the procedures is what to condition on when generating values for $\mu^t$
and $Q^t$. In the procedures used to generate the figures in the middle
row one is conditioning on all the ensemble members, but not the new
data. In the procedure for the upper row one is conditioning on the
new data and all the ensemble members except the ensemble member that
is to be updated. Other simulation runs not included in this report 
show that most of the difference in the results comes from not conditioning on the
ensemble member that is to be updated. The effect of including the
new data in the conditioning set is clearly visible, but still small
compared to the effect of not conditioning on the ensemble member that
is to be updated.

\begin{figure*}
  \begin{center}
    \begin{tabular}{@{}c@{}c@{}}
      \includegraphics[width=0.35\textwidth]{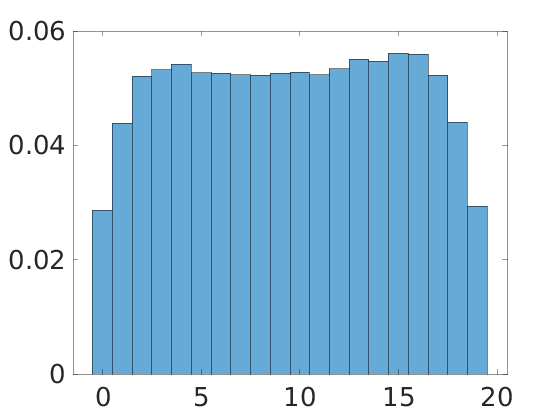}
      &
        \includegraphics[width=0.35\textwidth]{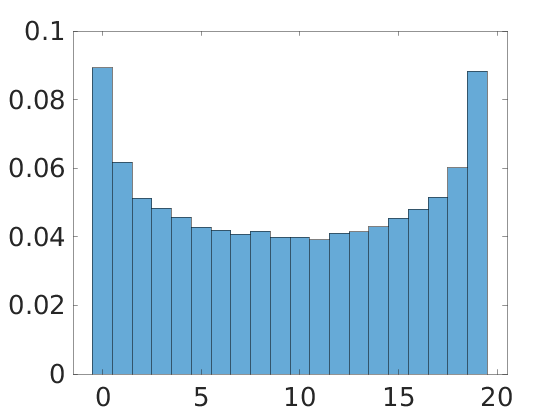}
      \\ \includegraphics[width=0.35\textwidth]{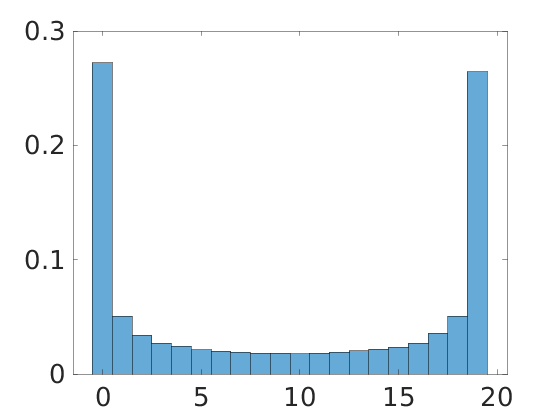}
      &
        \includegraphics[width=0.35\textwidth]{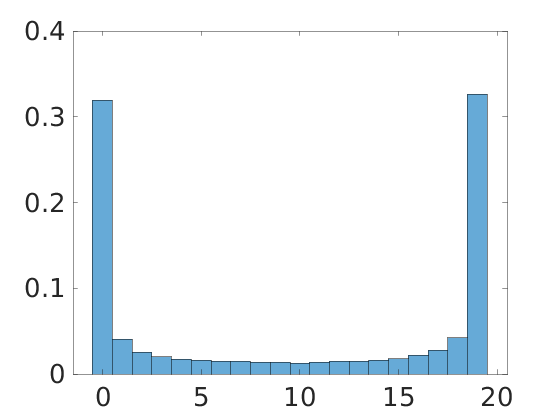}
      \\ \includegraphics[width=0.35\textwidth]{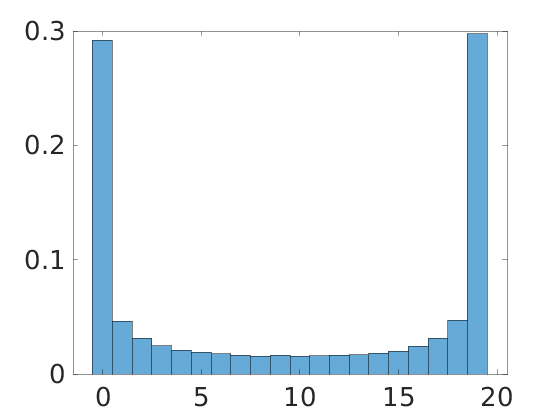}
      &
        \includegraphics[width=0.35\textwidth]{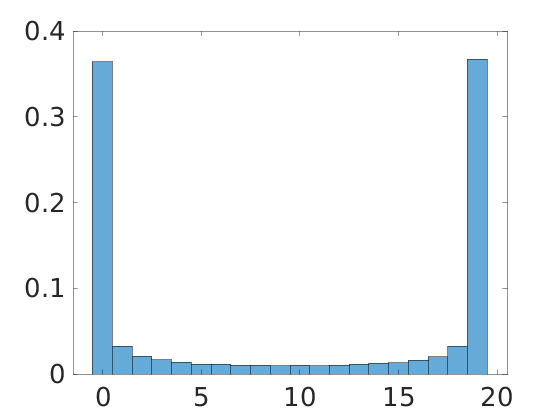}
      \\[-0.2cm]
    \end{tabular}
  \end{center}
  \caption{\label{fig:GaussianLinearHist}Gaussian linear example:
    Estimated distribution for $Z$ when using $M=19$ ensemble
    members. The upper, middle and lower rows are when using our proposed procedure
  for generating $\mu^t$ and $Q^t$, when using the procedure of
  \citet{art20} for the same, and when using empirical estimates,
  respectively. The left and right columns are when updating with our
  optimal square-root filter and when using the standard stochastic
  EnKF procedure, respectively.}
\end{figure*}

In the four lower plots in Figure \ref{fig:GaussianLinearState} the
latent true state vector is in most positions outside the spread of
the ensemble members. As such, these ensemble members do not give a
realistic representation of our information about $x^{11}$. In the two
upper plots in the same figure, the latent true state is in most
positions inside the spread of the ensemble members. These ensembles
may therefore give a better representation of the
uncertainty. However, the spread in the ensemble members is larger in
the upper left plot than in the upper right plot. So an interesting
question is therefore which of the two that gives the best
representation of our information about $x^{11}$. It is of course not
necessarily the procedure that gives the largest spread that gives the
best representation of uncertainty. To provide one answer to this
question, one can first observe that in a perfect model, 
the variables $x^{t, (1)},\ldots,x^{t, (M)},x^t$ are exchangeable. One
way to measure to what degree the spread of the ensemble members gives
a realistic representation of the uncertainty is therefore to study the
distribution of
\begin{equation}
  Z = \sum_{i=1}^M 1(x^{t, (i)}_j \leq x_j^t),
\end{equation}
where the index $j$ is sampled uniformly on the integers from $1$ to
$n$. In the perfect model
$Z$ has a uniform distribution on the integers zero
to $M$. Repeating the simulation procedures leading to the plots in
Figure \ref{fig:GaussianLinearState} one thousand times, randomising
also over the latent state vector, the plots in Figure
\ref{fig:GaussianLinearHist} show the estimated distributions for $Z$
for each of the six filtering procedures.
The four lower plots in this figure just confirm what we saw in Figure
\ref{fig:GaussianLinearState}, the latent state value is very often
more extreme than all the ensemble members. The distributions in the two upper
plots are neither perfectly uniform, but we see that the
distribution in the upper left plot is slightly closer to being
uniform than the upper right one. We
thereby conclude that of the six procedures tried here, it is our
proposed procedure that best represents our knowledge about $x^{11}$.

\begin{figure*}
  \begin{center}
    \begin{tabular}{@{}c@{}c@{}}
      \includegraphics[width=0.35\textwidth]{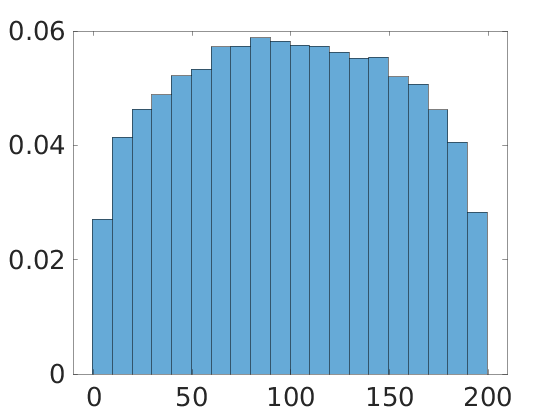}
      &
        \includegraphics[width=0.35\textwidth]{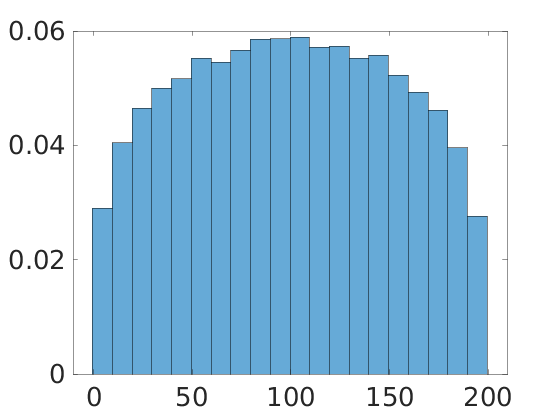}
      \\ \includegraphics[width=0.35\textwidth]{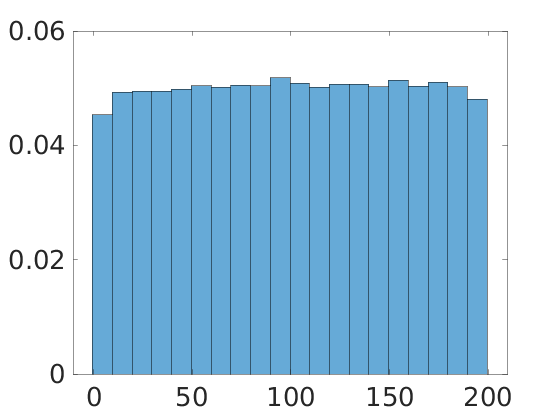}
      &
        \includegraphics[width=0.35\textwidth]{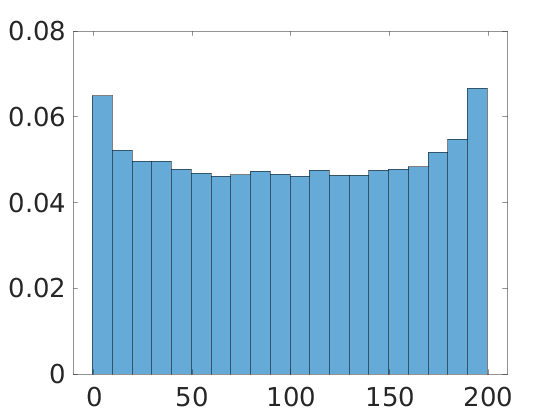}
      \\ \includegraphics[width=0.35\textwidth]{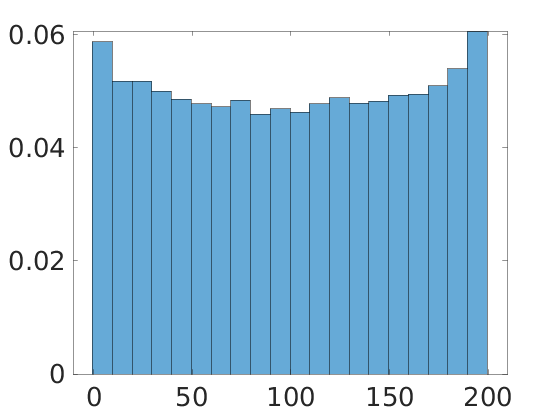}
      &
        \includegraphics[width=0.35\textwidth]{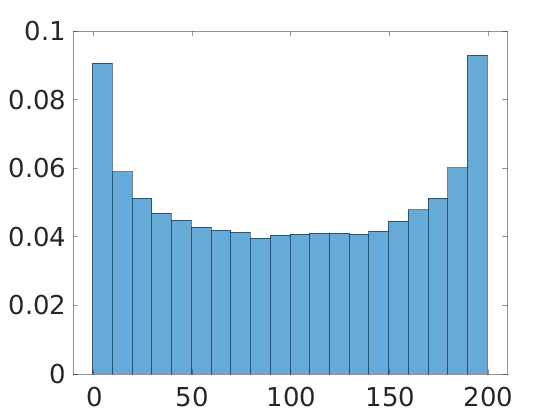}
      \\[-0.2cm]
    \end{tabular}
  \end{center}
  \caption{\label{fig:GaussianLinearHist199}Gaussian linear example:
    Estimated distribution for $Z$ when using $M=199$ ensemble
    members. The upper, middle and lower rows are when using our proposed procedure
  for generating $\mu^t$ and $Q^t$, when using the procedure of
  \citet{art20} for the same, and when using empirical estimates,
  respectively. The left and right columns are when updating with our
  optimal square-root filter and when using the standard stochastic
  EnKF procedure, respectively.}
\end{figure*}

Above,  we presented simulation experiments for the six ensemble updating procedures
we have defined, for a linear forward model and with $M=19$ ensemble
members. We have also done similar simulation experiments for both
smaller and larger ensemble sizes $M$, and for the non-linear forward function
discussed in Section \ref{sec:6.1}. There are two main
lessons to learn from these experiments. The first is that  the
differences between the six methods gradually reduce when the number
of ensemble members increases, and for $M$ large enough they all
behave essentially the same. It should, however, be remembered that in
typical applications of the EnKF,  the dimension of the state vector, $n$,  is much larger than the number of ensemble members,
$M$.  As one example, the plots in Figure
\ref{fig:GaussianLinearHist199} are the same type of plots as in Figure
\ref{fig:GaussianLinearHist}, but for runs with $M=199$ ensemble
members.

\begin{figure*}
  \begin{center}
    \begin{tabular}{@{}c@{}c@{}}
      \includegraphics[width=0.35\textwidth]{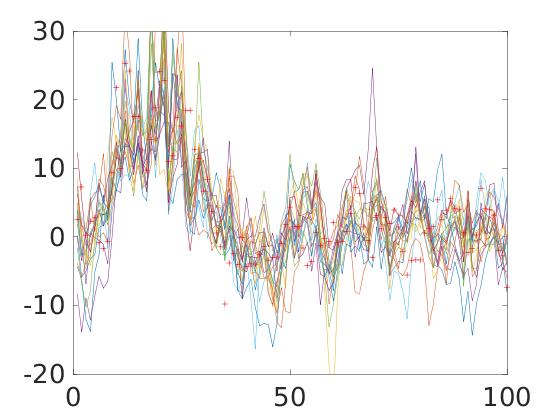}
      &
        \includegraphics[width=0.35\textwidth]{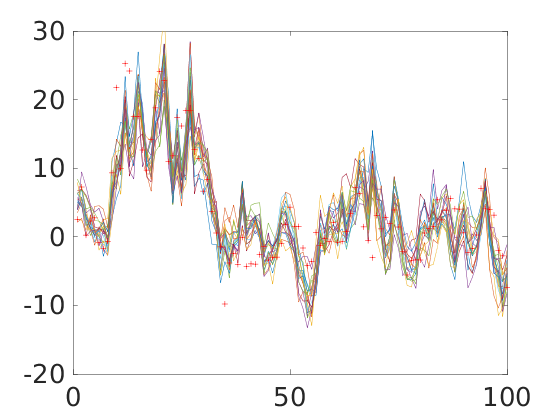}
      \\ \includegraphics[width=0.35\textwidth]{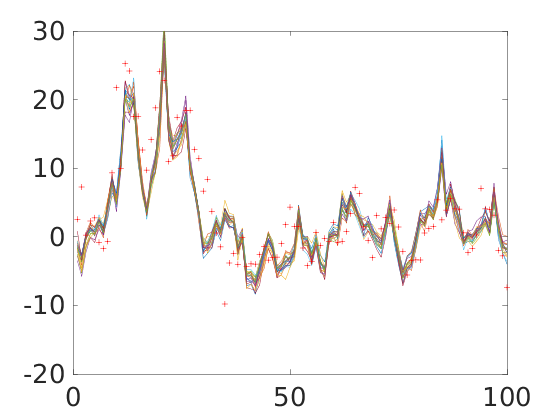}
      &
        \includegraphics[width=0.35\textwidth]{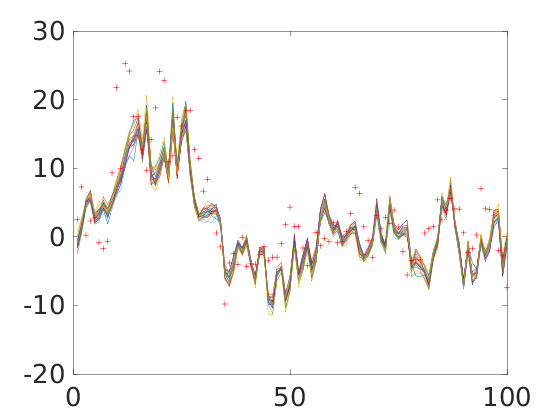}
      \\ \includegraphics[width=0.35\textwidth]{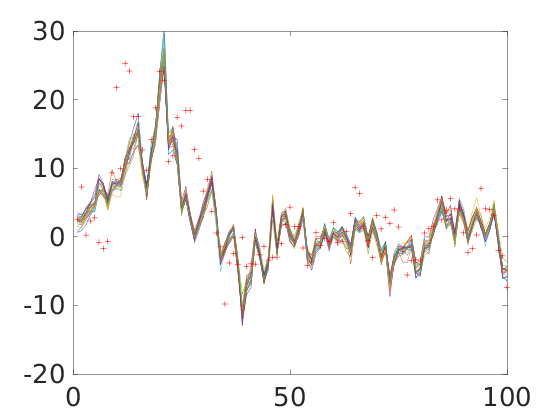}
      &
        \includegraphics[width=0.35\textwidth]{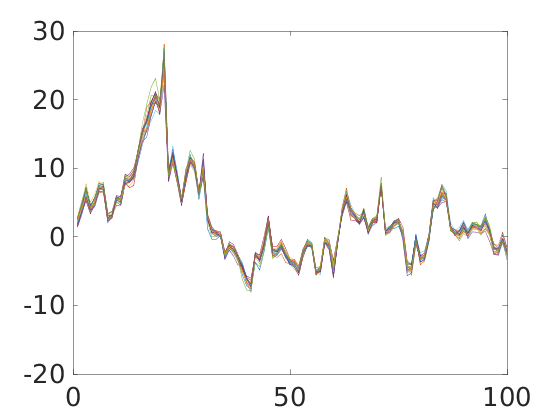}
      \\[-0.2cm]
    \end{tabular}
  \end{center}
  \caption{\label{fig:GaussianNonLinearState}Gaussian non-linear example:
    Prediction ensemble at time $T=11$ when using $M=19$ ensemble
    members. The upper, middle and lower rows are when using our proposed procedure
  for generating $\mu^t$ and $Q^t$, when using the procedure of
  \citet{art20} for the same, and when using empirical estimates,
  respectively. The left and right columns are when updating with our
  optimal square-root filter and when using the standard stochastic
  EnKF procedure, respectively. The ensemble members are shown with
  solid lines and the latent true state is shown with red crosses.}
\end{figure*}

\begin{figure*}
  \begin{center}
    \begin{tabular}{@{}c@{}c@{}}
      \includegraphics[width=0.35\textwidth]{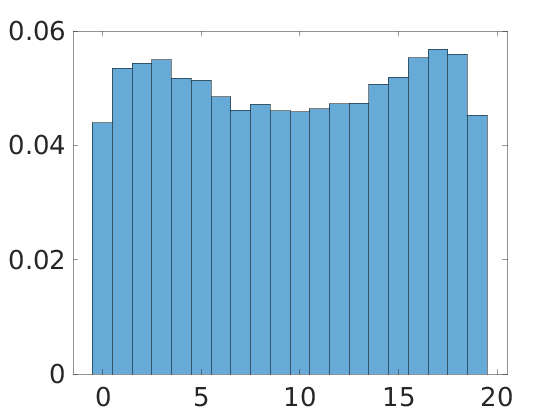}
      &
        \includegraphics[width=0.35\textwidth]{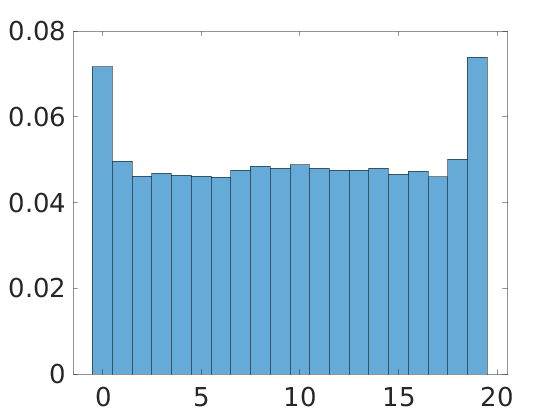}
      \\ \includegraphics[width=0.35\textwidth]{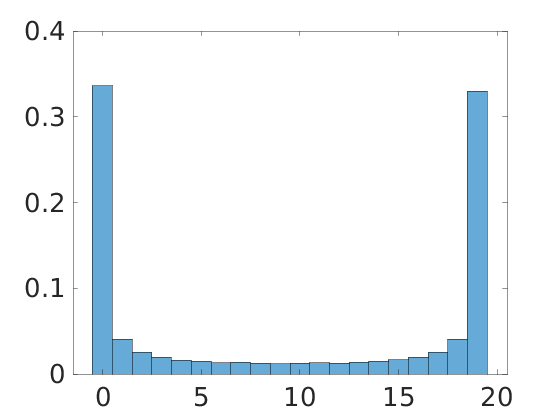}
      &
        \includegraphics[width=0.35\textwidth]{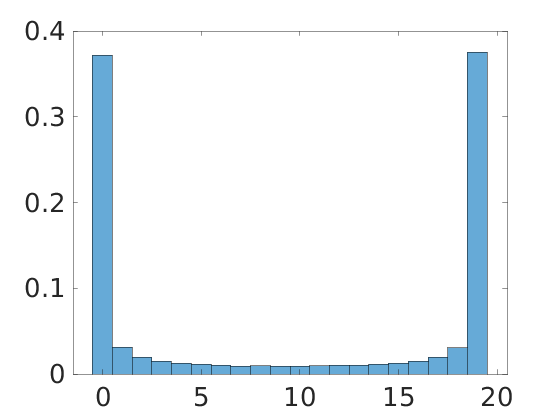}
      \\ \includegraphics[width=0.35\textwidth]{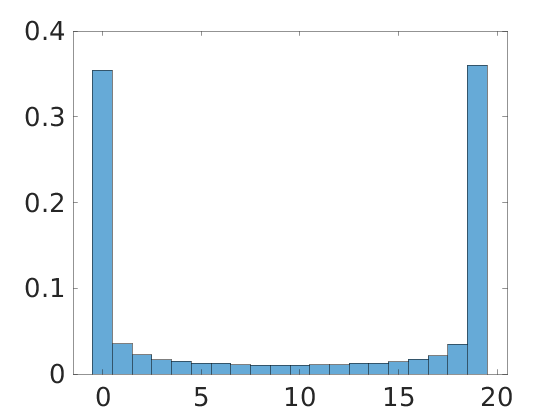}
      &
        \includegraphics[width=0.35\textwidth]{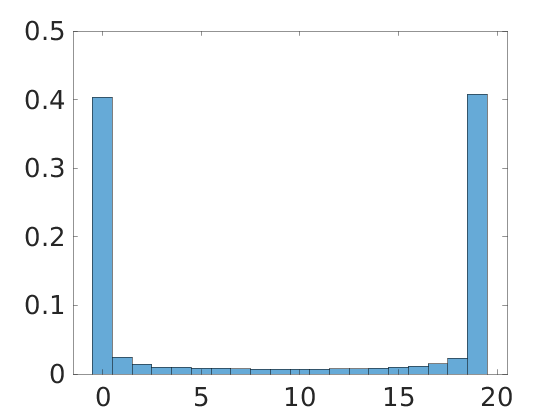}
      \\[-0.2cm]
    \end{tabular}
  \end{center}
  \caption{\label{fig:GaussianNonLinearHist}Gaussian non-linear example:
    Estimated distribution for $Z$ when using $M=19$ ensemble
    members. The upper, middle and lower rows are when using our proposed procedure
  for generating $\mu^t$ and $Q^t$, when using the procedure of
  \citet{art20} for the same, and when using empirical estimates,
  respectively. The left and right columns are when updating with our
  optimal square-root filter and when using the standard stochastic
  EnKF procedure, respectively.}
\end{figure*}

The second lesson we learn from the simulation experiments, is that
the results when using our 
non-linear forward function is quite similar to what we have for the
linear forward function. As one example, Figures \ref{fig:GaussianNonLinearState} and
\ref{fig:GaussianNonLinearHist}
show similar plots as in Figures
\ref{fig:GaussianLinearState} and \ref{fig:GaussianLinearHist}, but
for the non-linear forward function defined by Eqs. 
(\ref{eq:forwardNonlinear1}) and (\ref{eq:forwardNonlinear2}). Again
we see that the upper left plot in Figure
\ref{fig:GaussianNonLinearHist} is the one closest to being uniform.
Also
when using the non-linear forward function the differences between the
six methods gradually vanish when the number of ensemble members, $M$,
increases. Of course, that the results  for our non-linear
forward function are similar to the results for the linear function, does not imply that this is
generally true for all non-linear forward functions. We have for example not
studied how the various procedures perform with a forward function inducing skewed distributions for the state vector.

\section{Simulation experiment with a first-order Markov chain  assumed model}  
\label{sec:7}
In this section,  we demonstrate the proposed updating procedure in a simulation example where the state vector consists of binary variables and $f_{x^t | \theta^t}(x^t | \theta^t)$ and $f_{y^t | x^t}(y^t | x^t)$ constitute a hidden Markov model as described in Section \ref{sec:5}. The experimental setup of the simulation example is the same as in the simulation example presented in \cite{LoeTjelmeland2020}. 
Below,  we first describe the experimental setup of the simulation example  in Section \ref{sec:7.1}, and thereafter we 
present and discuss the simulation results  in Section \ref{sec:7.2}.

\subsection{Experimental setup}
\label{sec:7.1}
The simulation example involves a state process $\{x^t \}_{t=1}^T$  with $T=100$ time steps, and the state
vector $x^t$ at each time step is a vector of $n=400$ binary
variables, $x_j^t \in \{0,1\}$.
The initial distribution $p_{x^1}(x^1)$ and the forward model $p_{x^t | x^{t-1}}({x^t | x^{t-1}})$ of the unobserved $x^t$-process are the same as in the simulation example of \cite{LoeTjelmeland2020}. %
For simplicity,  we do not discuss the technical  details of this model here, 
but one should note that the generated state vector $x^t$ at any time $t$ is not a first-order 
Markov chain. 
The process is inspired by how water comes through to an oil-producing well in a petroleum reservoir. 
In this context, we let the $t$ in $x_j^t$ represent time and $j$ the location in the well, and the values zero and one represent oil and water, respectively. Hence, the event $x_j^t = 0$ indicates the presence of oil in location $j$ at time $t$, while the event $x_j^t = 1$ indicates the presence of water.

\begin{figure*}
\centering
\begin{tikzpicture}
\node[inner sep=0pt,label=below:\footnotesize (a)] (russell) at (0,0) {\includegraphics[width=.20\textwidth]{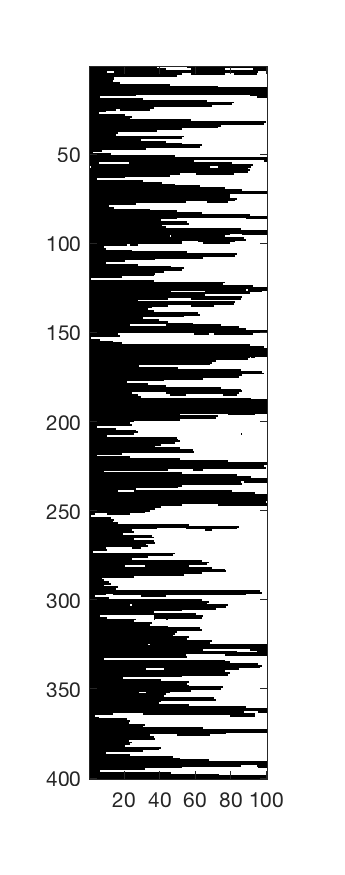}};
\node[inner sep=0pt,label=below:\footnotesize (b)] (n6) at (3,0) {\includegraphics[width=0.20\textwidth]{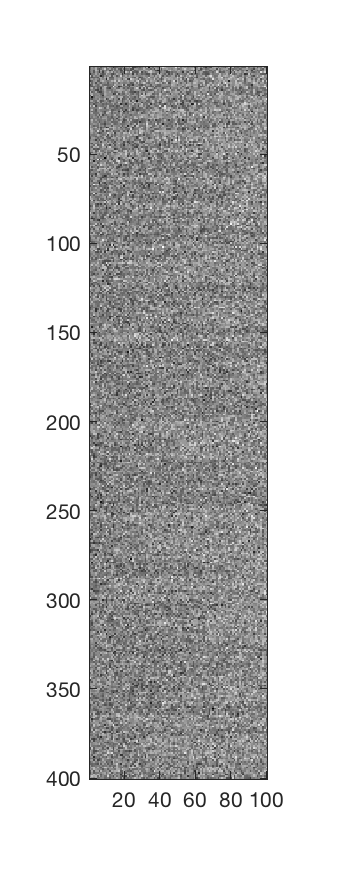}};
\node[inner sep=0pt,label=below:\footnotesize (c)] (n7) at (6,0) {\includegraphics[width=0.20\textwidth]{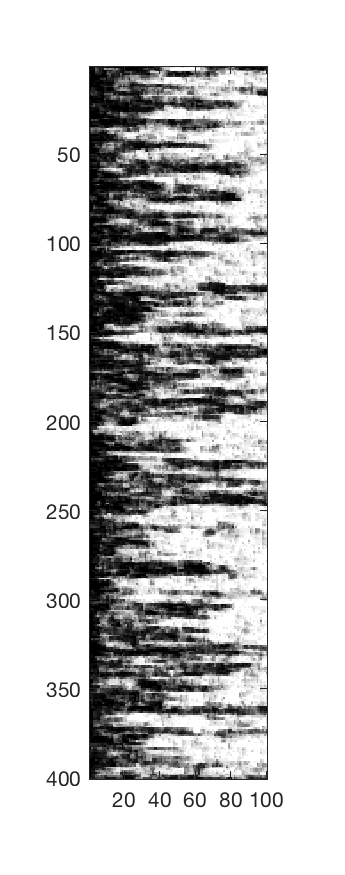}};
\node[inner sep=0pt,label=below:\footnotesize (d)] (n8) at (9,0) {\includegraphics[width=0.20\textwidth]{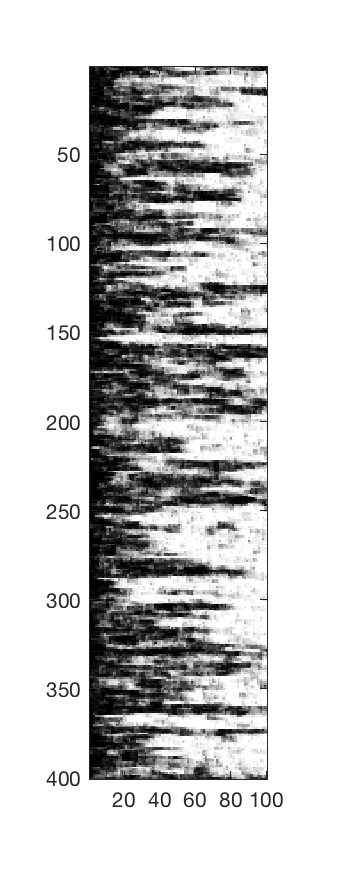}};
\node[draw=none,fill=none] (n0) at (-2,3.5) {};
\node[draw=none,fill=none] (n1) at (-2,2.5) {} edge[<-] (n0);
\node[draw=none,fill=none] (n2) at (-1,3.5) {} edge[<-] (n0);
\node [draw=none,fill=none] (n3) at (-2.25,3) {$j$};
\node [draw=none,fill=none] (n4) at (-1.5,3.75) {$t$};
\end{tikzpicture}
\caption{First-order Markov chain simulation example: (a)  the latent state process, (b) the observations, (c) estimates of marginal filtering probabilities obtained with the proposed Bayesian updating approach, and (d)  estimates of marginal filtering probabilities obtained with the non-Bayesian updating approach. In all figures, the colour black represents the value zero and the colour white represent the value one. }
\label{fig:1}

\end{figure*}
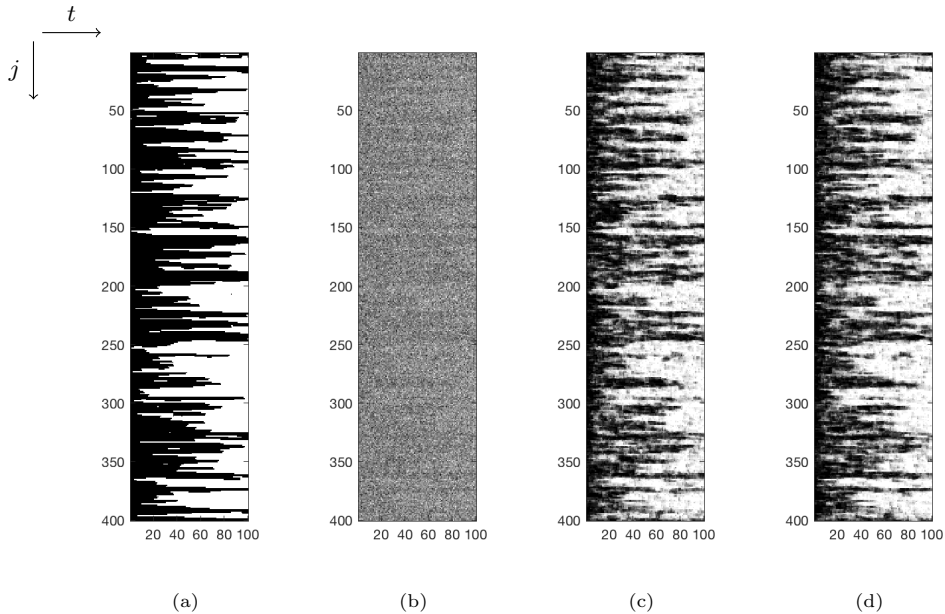

An image of a state process $\{x^t \}_{t=1}^T$ generated using the true model specified above is shown in Figure \ref{fig:1}(a), where the colours black and white represent the values zero (oil) and one (water), respectively. 
Based on this reference state process,   a corresponding observation process $\{y^t\}_{t=1}^T$ is generated by simulating,  independently 
for each time step $t=1,\ldots,T$ and for each node $j=1,\ldots,n $,
an observation $y^t_j$ from a Gaussian distribution with mean $x^t_j$
and variance $\sigma^2=2^2$.
Figure \ref{fig:1}(b) shows a grey-scale image of the generated observation process. 
Pretending that only the observations   are available, the goal is to assess the  filtering distribution $p_{x^t | y^{1:t}}(x^t | y^{1:t})$ for each time step $t=1, \dots, T$.

As described in Section \ref{sec:5}, the assumed model $f_{x^t | \theta^t}(x^t | \theta^t)$ is a first-order Markov chain, and the parameter  $\theta^t $ represents its initial and transition probabilities. Moreover, $\theta^t$ is a vector of the Dirichlet distributed random variables
$$
  \theta_1^t = (\theta_1^t(0), \theta_1^t(1)), 
$$
$$
  \theta_j^{t,0} = (\theta_j^{t,0}(0), \theta_j^{t,0}(1)), 
$$
  and  
$$
  \theta_j^{t,1} = (\theta_j^{t,1}(0), \theta_j^{t,1}(1)), 
$$
for $j = 2, \dots, n$. 
The corresponding hyperparameters $\alpha_1^t(0)$, $\alpha_1^t(1)$, $\alpha_j^{t,0}(0)$,  $\alpha_j^{t,0}(1)$,
 $\alpha_j^{t,1}(0)$,  $\alpha_j^{t,1}(1)$  are all set equal to 2 at every time step $t$. 
 For the assumed likelihood $f_{y^t | ^t}(y^t  | x^t)$ we  use the same distribution as the one
used to simulate the data; that is,  each distribution $f_{y_j^t | x_j^t}(y_j^t | x_j^t)$ is a Gaussian with mean $x_j^t$ and variance $\sigma^2 = 2^2$. In the Gibbs simulation of $\theta^{t,(i)} | x^{t,-(i)}, y^t$,   100 iterations are used. Finally, as  in \cite{LoeTjelmeland2020}, we use the ensemble size $M=20$.  

\subsection{Simulation results}
\label{sec:7.2}
To evaluate the performance of the proposed approach, we compare our results  with corresponding results obtained using the method of \cite{LoeTjelmeland2020}. 
For simplicity,  we refer in the following to the method proposed in the present report as the Bayesian approach, and the method proposed in \cite{LoeTjelmeland2020} as the non-Bayesian approach.

\begin{figure}[t]
\centering
\subfigure{\includegraphics[width=0.65\textwidth]{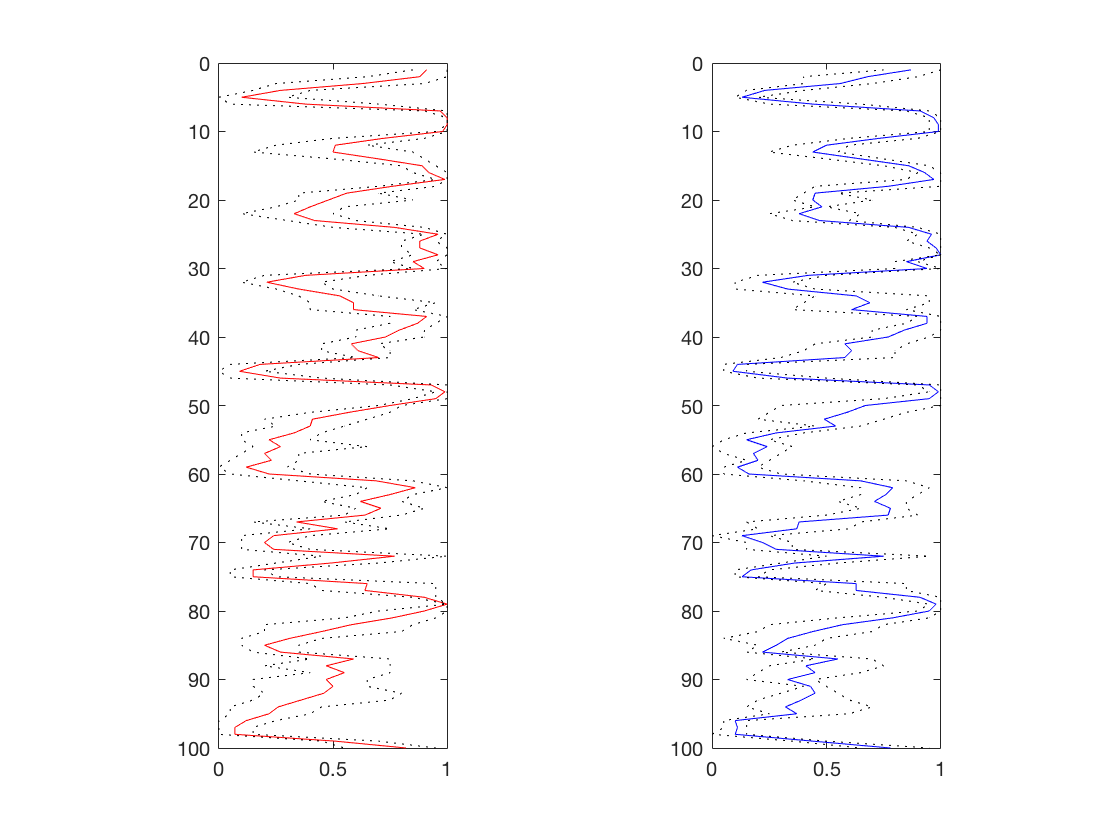}}
\caption{First-order Markov chain simulation example: The left plot shows the empirical means (solid red line) of five estimated values $\hat p(x_j^t = 1| y^{1:t})$ for the marginal filtering probability $p_{x_j^t | y^{1:t}}(x_j^t = 1 | y^{1:t})$ obtained from five independent runs of the Bayesian approach, along with the corresponding minimum and maximum values (dotted black lines) of the five estimates. The right plot shows corresponding output from the non-Bayesian approach.  }
\label{fig: marginal est}
\end{figure}

Figures \ref{fig:1}(c) and (d) show grey-scale images of estimated  values $\hat p(x_j^t = 1 | y^{1:t})$ of the marginal filtering probabilities $p_{x_j^t | y^{1:t}}(x_j^t = 1 | y^{1:t})$, $j=1, \dots, n$, $t=1, \dots, T$ obtained with the Bayesian and the non-Bayesian approach, respectively, where the estimate $\hat p(x_j^t = 1 | y^{1:t})$ is  the empirical mean of the $\tilde x^{t,(i)}_j$-samples, 
\begin{equation}
\hat p(x_j^t = 1 | y^{1:t}) = \frac{1}{M} \sum_{i=1}^{M} \tilde x_j^{t,(i)}. 
\label{eq: marginal filt}
\end{equation}
From a visual inspection, the output from the two approaches look very similar. 
To investigate this further,  we perform five independent runs of each method and estimate the marginal filtering probabilities in each run. For each of the two methods, we thereby obtain five samples, $\hat p^{(r)}(x_j^t = 1 | y^{1:t})$, $r=1, \dots, 5$,  of $\hat p(x_j^t = 1 | y^{1:t})$ in  Eq. \eqref{eq: marginal filt}. 
Figure \ref{fig: marginal est} 
shows plots of  the empirical means of these five samples for locations $j=1$ to 100 at the (arbitrarily chosen) time step $t=50$, along with the corresponding minimum and maximum values of the five samples. 
Equivalent output from other time steps $t$ and for other locations $j$ follow the same trend and are therefore, for simplicity, not included. 
As seen in Figure \ref{fig: marginal est}, the results from the two methods look very much the same. 
This may suggest 
that the Bayesian approach offers no considerable advantage over the non-Bayesian approach, at least not  when it comes to estimating marginal filtering probabilities. 

\begin{figure}[t]
\centering
\subfigure{\includegraphics[width=0.65\textwidth]{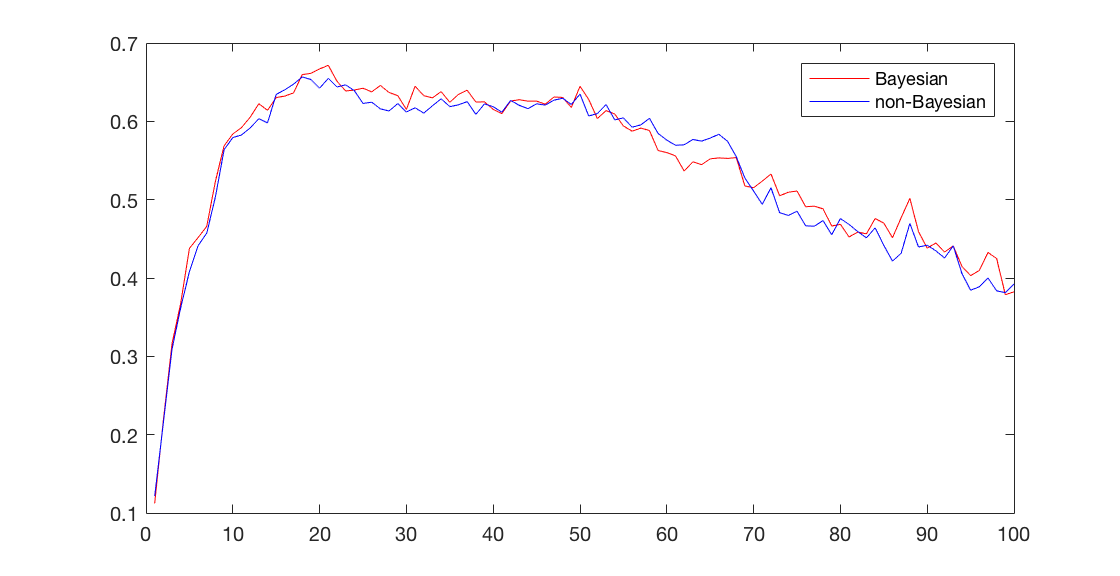}}
\caption{First-order Markov chain simulation example: Plots of the approximated coefficients of unalikeability, $\bar u^t$, computed at each time step $t=1,\dots, 100$, for the Bayesian approach (red) and the non-Bayesian approach (blue). }
\label{fig:variability}
\end{figure}

Methodologically, the main difference between the Bayesian and the non-Bayesian approach is that $\theta^t$ is treated as random in the Bayesian approach. More specifically, the Bayesian approach simulates a parameter value $\theta^{t,(i)}$ for each ensemble member 
$x^{t,(i)}$, while the non-Bayesian approach instead computes an estimate, $\hat \theta^t$, and this same estimate $\hat \theta^t$ is used to update  all the forecast samples. 
Therefore,  since the Bayesian approach incorporates randomness in $\theta^t$, one would expect the spread, or the variability, in the  samples  from the 
Bayesian approach to be greater than the variability in the samples  from the non-Bayesian approach,
which is also what \cite{art20} observed in their work and what we observed in the simulation example with  the linear-Gaussian model presented in the previous section.
However,  it appears that this is not the case for the binary simulation experiment  studied here. 
For continuous variables, variability is easy to measure and visualise, but for categorical variables,  other techniques are necessary. 
To study the variability of the results in the categorical context of this example,  we consider the \emph{coefficient of unalikeability} (CU) of  \cite{art27}. 
Given a set of independent random samples taking values in a categorical sample space, the CU provides a measure for how unalike the samples are. 
In the present simulation example, we are interested in computing the CU of the filtering ensemble $\{ \tilde x^{t,(1)}, \dots,\tilde x^{t,(M)} \}$ at each of the time steps  $t = 1, \dots, T$. Hereafter, we denote the CU of $\{ \tilde x^{t,(1)}, \dots,\tilde x^{t,(M)} \}$ by $u^t$. 
Since $\tilde x^{t,(i)}$ is a vector of  $n=400$ binary variables, there are  $2^{400}$ possible configurations for $\tilde x^{t,(i)}$. Each configuration can be interpreted as a (unique) category.  Hence, each realisation  $\tilde x^{t,(i)}$ of the posterior ensemble corresponds to one of the $2^{400}$ possible categories. 
However, we only have $M=20$ ensemble members, which is not enough to give an informative value for $u^t$ when the number of categories is so high. 
Therefore, we consider first each  four-tuple $x_{j:j+3}^t = (x^t_j, x^t_{j+1}, x^t_{j+2}, x^t_{j+3})$, $j=1, \dots, n-3$, of $x^t$ separately. The number of possible configurations for each such four-tuple is $2^4 = 16$,  and from the  posterior samples $\tilde x_{j:j+3}^{t,(1)}, \dots, \tilde x_{j:j+3}^{t,(M)}$ we can compute a coefficient of unalikeability $u_j^t$. 
After having computed $u^t_j$ for each four-tuple $x_{j:j+3}^t $ of $x^t$, we compute the mean, $\bar u^t$, of all of them. This $\bar u^t$ then serves  as an approximation for the actual CU,  $u^t$,  of  $\{ \tilde x^{t,(1)}, \dots,\tilde x^{t,(M)} \}$.  
Figure \ref{fig:variability} shows a plot of the values of $\bar u^t$, $t=1, \dots, T$, obtained with the Bayesian approach (red line) and the non-Bayesian approach (blue line). 
As one can see, the values of $\bar u^t$ from the Bayesian approach very much coincide with the values from the non-Bayesian approach,  which indicates a similar variability in the samples.

After various additional tests, both with different data $\{y^t\}_{t=1}^T$, different values for the observation noise $\sigma$ and different values for the ensemble size $M$, 
it seems that the variability in the results from the two approaches, and the results from the two approaches in general, are very much alike. 
One possible reason for this, is the optimality criterion for $q(\tilde x^{t,(i)} | x^{t,(i)}, \theta^t, y^t)$, i.e. the criterion of maximising  the expected number of unchanged components of $x^{t,(i)}$. 
Basically, the optimality criterion states that we want to make
minimal changes to the forecast samples, and this results in that the distributions 
$q(\tilde x^{t,(i)} | x^{t,(i)}, \theta^{t,(i)}, y^{t})$, $i=1, \dots, M$, from
the Bayesian approach and the distribution $q(\tilde x^{t,(i)} | x^{t,(i)}, \hat \theta^{t}, y^t)$
from  the non-Bayesian approach are all drawn towards each other. 
Consequently, the generated posterior samples from the two approaches
will be similar to each other.  
Another possible reason for the lack of differing variability is the binary nature of the problem. More specifically, since 
both approaches capture the mean of $x_j^t$ quite well, they must also capture the variance, as there is a one-to-one relationship between the mean and variance for a binary random variable.

\section{Closing remarks}
\label{sec:8}

In this report,  a general framework for updating a prior ensemble to a
posterior ensemble is presented. Being able to update a prior ensemble
to a posterior ensemble is a crucial step in ensemble-based solutions
to the filtering, or data assimilation, problem.  
The proposed method is based on an assumed Bayesian model and a proposed 
optimality criterion.

Two special applications of the general framework are investigated,
one where the elements of the state vector are continuous variables
and one where the elements are binary  variables.  
In the continuous case, an assumed Gaussian distribution is adopted
for the state vector and a linear-Gaussian model for the observation.
This results in a class of updating methods where a fully Bayesian version of the EnKF is 
a special case, and we prove that a particular version of the square
root EnKF is optimal with respect to the optimality criterion of
making minimal changes to each ensemble member.  
In the binary 
application, the state and observation vectors are instead assumed to
follow a finite state-space HMM.  The corresponding updating
procedure is then essentially the same as the one for binary vectors
proposed in \cite{LoeTjelmeland2020},  except now the transition probabilities of
the assumed Markov chain model are treated as random.

When studying the results of the presented simulation examples, the
most striking aspect is that the proposed approach is substantially better
in representing the uncertainty in the situation with the  linear-Gaus\-sian
model. When comparing results from the proposed approach with
results obtained using the procedure of \cite{art20},  we really see
the importance of not using the same information twice. That we do not
get the same dramatic effect in the example for the assumed HMM
may be because in that model the same parameters control both the
mean and the variance. As the non-Bayesian ensemble filtering method seems
to capture the mean quite well, it must then 
also give a good representation of the variance.

Computational efficiency is not a main focus in the present
report. The dynamic programming procedure developed for the assumed
HMM requires computing time proportional to the number of elements in
the state vector and is thereby computationally efficient. The
updating procedure of the assumed linear-Gaussian model requires
inversion of $n\times n$ matrices, where $n$ is the dimension of the
state vector, so this procedure is only computationally feasible for
sufficiently small values of $n$. In typical applications of the EnKF,  the
state vector is very large and computational efficiency is therefore
essential. In the EnKF,  the prior covariance matrix is estimated by the
empirical covariance matrix of the prior ensemble. The rank of the
(estimated) covariance matrix is thereby limited by the number of
ensemble members, which is typically much smaller than the dimension
of the state vector. The low rank of the covariance matrix makes it
possible to rephrase the EnKF updating equation so that efficient
computation is possible.  In the proposed approach for the assumed
linear-Gaussian model, the generated covariance matrices
are by construction of full rank. It should, however, be possible to
get computational efficiency by restricting the inverse covariance
matrices, i.e. precision matrices, to be sparse. To achieve this, a prior tailored
to produce sparse precision matrices must be constructed and the class
of updating distributions must be restricted to ensure that all
necessary computations for the updating can be performed on sparse
matrices. The details of this is a direction of future research.

In the present report, we have studied in detail two applications
of the proposed framework. In the future,  it is of interest to explore
also other assumed models and other optimality criteria. It would in
particular be interesting to consider a situation where the state
vector represents a two-dimensional lattice of categorical
variables. A possible assumed prior model is then a Markov
mesh model \citep{Abend1965}. It would also be interesting to apply the proposed framework in a
mixed discrete and continuous situation, i.e. a model where the state
vector consists of both discrete and continuous variables.  

\bibliographystyle{apa}
\bibliography{biblio}

\appendix

\section{Proof of the result in Example \ref{example3}} 
\label{sec:A2}

Here we prove the result stated in Example \ref{example3}; that is, we prove that  when $B^t$ and $S^t$ are as specified in Eqs. \eqref{eq:24} and \eqref{eq:25}, respectively, the linear update in Eq. \eqref{eq:23} corresponds to the stochastic EnKF update in Eq. \eqref{eq:stochEnKF}.

We start by inserting the expression for $B^t$ in
Eq. \eqref{eq:24}
into Eq. \eqref{eq:23}. This gives
\begin{equation}
  \tilde x^{t, (i)} = x^{t, (i)} + K^t(y^t - H^tx^{t, (i)}) + \tilde \epsilon^{t, (i)}.
  \label{eq:a1}
\end{equation}
Comparing Eq. \eqref{eq:a1} with the stochastic EnKF update in  Eq. \eqref{eq:stochEnKF} we see that it remains to
show that the distribution of $\tilde \epsilon^{t, (i)}$ in 
Eq. \eqref{eq:a1} is identical to the distribution of $K^t\epsilon^{t, (i)}$
in Eq. \eqref{eq:stochEnKF}. As both $\tilde \epsilon^{t, (i)}$ and
$\epsilon^{t, (i)}$ are Gaussian with zero mean, the distributions of
$\tilde \epsilon^{t, (i)}$ and $K^t\epsilon^{t, (i)}$ are equal if
\[
  \mbox{Cov}\!\left[\tilde \epsilon^{t, (i)}\right] =
  \mbox{Cov}\!\left[K^t \epsilon^{t, (i)}\right].
\]
Since we have $\text{Cov}[\tilde \epsilon^{t, (i)}]  = S^t$, with $S^t$ given by Eq.  \eqref{eq:25}, 
and $ \text{Cov} [K^t \epsilon^{t, (i)}]= K^tR^t(K^t)^\top$, this means that we need to show that
\begin{equation*}
(I_n-K^tH^t)Q^t(K^t(H^t)^\top) = K^tR^t(K^t)^\top,
\end{equation*}
or rather
\begin{equation}
(I_n-K^tH^t)Q^t(H^t)^\top = K^tR^t.
\label{eq: want to prove this}
\end{equation} 
In order to prove Eq.  \eqref{eq: want to prove this} we first prove that
\begin{equation}
\left ( (Q^t)^{-1} + (H^t)^\top (R^t)^{-1} H^t \right )^{-1} (Q^t)^{-1} = I_n-K^tH^t
\label{eq: prove1}
\end{equation}
and
\begin{equation}
\left ( (Q^t)^{-1} + (H^t)^\top (R^t)^{-1} H^t \right )^{-1} (H^t)^\top (R^t)^{-1} = K^t.
\label{eq: prove2}
\end{equation}
To prove Eqs.  \eqref{eq: prove1} and \eqref{eq: prove2} we make use of the following two formulations of the Woodbury matrix identity,
\begin{align}
  \left ((Q^t)^{-1} + (H^t)^\top (R^t)^{-1} H^t \right )^{-1} 
  = Q^t + Q^t(H^t)^\top \left (R^t+H^tQ^t(H^t)^\top \right )^{-1} H^tQ^t,
                \label{eq: wood1}
   \end{align}
   \begin{align}
  \left ( R^t + H^tQ^t(H^t)^\top  \right )^{-1} 
  = (R^t)^{-1} - (R^t)^{-1} H^t \left ( (Q^t)^{-1} + (H^t)^\top (R^t)^{-1} H^t \right )^{-1} (H^t)^\top (R^t)^{-1}. 
\label{eq: wood2}
\end{align}
To prove Eq.  \eqref{eq: prove1} we start by inserting Eq. \eqref{eq: wood1} on
the left hand side in Eq.  \eqref{eq: prove1}
and use that the Kalman gain is given as  $K^t = Q^t(H^t)^\top \left ( H^tQ^t(H^t)^\top + R^t \right )^{-1}$, 
\begin{eqnarray*}
  \left (  (Q^t)^{-1} + (H^t)^\top (R^t)^{-1} H^t \right )^{-1} (Q^t)^{-1}
  &=& \left ( Q^t + Q^t (H^t)^\top \left (R^t+H^tQ^t (H^t)^\top \right )^{-1} H^tQ^t \right ) (Q^t)^{-1}  \\
   &=& I_n - K^tH^t. 
\end{eqnarray*}
Hence we see that the left hand side and the right hand side in Eq.  \eqref{eq: prove1} are equal, and Eq. \eqref{eq: prove1} is thereby proved.
To prove Eq.  \eqref{eq: prove2} we start by considering
$\left ( (Q^t)^{-1} + (H^t)^\top (R^t)^{-1} H^t \right ) K^t$, insert that $K^t = Q^t(H^t)^\top \left ( H^tQ^t(H^t)^\top + R^t \right )^{-1}$, 
and use the Woodbury identity in Eq. \eqref{eq: wood2}. Specifically,
\begin{align*}
&\left ( (Q^t) ^{-1} + (H^t)^\top (R^t)^{-1} H^t \right )   K^t 
= \left ( (Q^t)^{-1} + (H^t)^\top (R^t)^{-1} H^t \right )  Q^t(H^t)^\top \left (H^tQ^t(H^t)^\top + R^t \right )^{-1} \\
  &= \left ( (Q^t)^{-1} + (H^t)^\top (R^t)^{-1} H^t \right )  Q^t(H^t)^\top 
    \left ((R^t)^{-1} - (R^t)^{-1}H^t \left ( (Q^t)^{-1}+(H^t)^\top(R^t)^{-1}H^t \right )^{-1}(R^t)^{-1}  \right ) \\
  &=  \left ( (H^t)^\top + (H^t)^\top(R^t)^{-1}H^tQ^t(H^t)^\top   \right )
   \left ( (R^t)^{-1} - (R^t)^{-1}H^t( (Q^t)^{-1}+(H^t)^\top(R^t)^{-1}H^t )^{-1}(R^t)^{-1}  \right  )\\
  &= (H^t)^\top (R^t)^{-1} + (H^t)^\top (R^t)^{-1} H^tQ^t(H^t)^\top (R^t)^{-1}  
  \\ &~~~~ - (H^t)^\top (R^t)^{-1} H^t \left ( (Q^t)^{-1} + (H^t)^\top (R^t)^{-1} H^t \right )^{-1} (H^t)^\top (R^t)^{-1}  \\ 
  &~~~~ - (H^t)^\top (R^t)^{-1} H^tQ^t (H^t)^\top (R^t)^{-1} H^t \left ( (Q^t)^{-1} + (H^t)^\top (R^t)^{-1} H^t\right )^{-1} (H^t)^\top (R^t)^{-1}
\\
  & = (H^t)^\top (R^t)^{-1} - (H^t)^\top (R^t)^{-1} H^tQ^t \\
  &~~~~\cdot \left [ -I_n + \left ( (Q^t)^{-1} + (H^t)^\top (R^t)^{-1} H^t \right )  \left ( (Q^t)^{-1} + (H^t)^\top (R^t)^{-1} H^t \right )^{-1} \right ] (H^t)^\top (R^t)^{-1}\\ 
& = (H^t)^\top (R^t)^{-1}.
\end{align*}
Hence we have shown that
\[
 \left ( (Q^t)^{-1} +  (H^t)^\top (R^t)^{-1} H^t \right )   K^t = (H^t)^\top (R^t)^{-1}.
\]   
Multiplying by  $\left ( (Q^t)^{-1} +  (H^t)^\top (R^t)^{-1} H^t \right ) ^{-1}$ on both sides, we get Eq. \eqref{eq: prove2}. 
Now, to prove Eq. \eqref{eq: want to prove this} we insert  Eq. \eqref{eq:
  prove1} on the left hand side of Eq. \eqref{eq: want to prove this}  
and  insert Eq. \eqref{eq: prove2} on the right hand side of Eq.  \eqref{eq: want to prove this}. Specifically,  
the left hand side of Eq. \eqref{eq: want to prove this} then reads
\begin{eqnarray}
(I_n-K^tH^t)Q^t(H^t)^\top = \left ((Q^t)^{-1} + (H^t)^\top (R^t)^{-1} H^t \right )^{-1} (Q^t)^{-1} Q^t (H^t)^\top \\ = \left ( (Q^t)^{-1} + (H^t)^\top (R^t)^{-1} H^t \right )^{-1}  (H^t)^\top, 
\label{eq: final1}
\end{eqnarray}
while the right hand side reads 
\begin{align}
  \left ( (Q^t)^{-1} + (H^t)^\top (R^t)^{-1} H^t \right )^{-1} (H^t)^\top (R^t)^{-1} R^t
  = \left ( (Q^t)^{-1} + (H^t)^\top (R^t)^{-1} H^t \right )^{-1} (H^t)^\top.
\label{eq: final2}
\end{align}
We see that Eqs. \eqref{eq: final1} and \eqref{eq: final2} are equal, and the proof is complete.

\section{Proof of Theorem \ref{thm}}

\label{sec:A1}

For any real matrices $M$ and $N$ of equal  dimension, let $\langle M,N \rangle $ denote the Frobenius inner product,
\[
\langle M,N \rangle = \text{tr}(MN^\top)
\]
The Cauchy-Schwarz inequality, $| \langle M,N \rangle |^2 \leq \langle M,M \rangle \langle N,N \rangle $,  then gives  
\[
\text{tr} (MN^\top) ^2 \leq \text{tr}(MM^\top) \text{tr}(NN^\top)
\]
with equality if and only if there exists a constant $c \in \mathbb R$ such that $M=cN$.

Using the singular value decomposition of $Z$, i.e. $Z = PGF^\top$, we can write
\begin{align}
\text{tr}(\tilde B Z) 
= \text{tr}( \tilde B  P G F^\top) 
= \text{tr}( \tilde B P G^{\frac{1}{2}} (FG^{\frac{1}{2}} )^\top ). 
\end{align}
The Cauchy-Schwarz inequality for $ \text{tr} \left ( \tilde B P G^{\frac{1}{2}} (FG^{\frac{1}{2}} )^\top  \right )$ 
with $M = \tilde B P G^{\frac{1}{2}} $ and $N=FG^{\frac{1}{2}}$  then gives
\begin{eqnarray}
\text{tr} \left (\tilde B Z \right ) ^2 
&\leq& \text{  tr} \left  (\tilde B P G^{\frac{1}{2}}  (\tilde B P G^{\frac{1}{2}})^\top \right  ) \text{tr} \left  ( FG^{\frac{1}{2}} (FG^{\frac{1}{2}} )^\top \right  )
\label{eq: CS}
\end{eqnarray}
with equality if and only if there exists a number $c \in \mathbb R$ such that 
\[
\tilde B PG^{\frac{1}{2}} = c F G^{\frac{1}{2}} \iff \tilde B = cFP^\top. 
\]
Using basic trace properties and that  $\tilde B^\top \tilde B = I_n-\tilde S$ and $F^\top F = P^\top P = I_n$, 
the right hand side in \eqref{eq: CS} can be rewritten as
\begin{align*}
&\text{  tr} \left  (\tilde B P G^{\frac{1}{2}}  (\tilde B P G^{\frac{1}{2}})^\top \right  ) \text{tr} \left  ( FG^{\frac{1}{2}} (FG^{\frac{1}{2}} )^\top \right  )\\
&~~~~= \text{tr} \left (\tilde BPG P^\top \tilde B^\top \right ) \text{tr} \left (F G  F^\top \right ) 
\\
&~~~~= \text{tr} \left ( PG P^\top \tilde B^\top \tilde B    \right ) \text{tr} \left ( GF^\top F \right ) 
\\
&~~~~= \text{tr} \left (   PGP^\top (I-\tilde S) \right ) \text{tr} \left ( G \right )
\\
&~~~~= \left ( \text{tr} \left ( PGP^\top \right ) - \text{tr} \left ( PGP^\top \tilde S   \right ) \right ) \text{tr} (G)
\\
&~~~~= \left ( \text{tr}(G)  - \text{tr}(PGP^\top \tilde S)   \right ) \text{tr} (G).
\end{align*}
When $\tilde S = 0$, we see that the Cauchy-Schwarz inequality yields
\[
\text{tr}(\tilde B Z)^2 \leq \text{tr}(G)^2
\]
with equality if and only if there exists $c \in \mathbb R$ such that $\tilde B = c FP^\top$. 
The condition that $\tilde S = I_n-\tilde B^\top \tilde B=0$ gives restrictions on the allowed values for $c$. 
Specifically,
\begin{align*}
I_n - \tilde B^\top \tilde B &= I_n - (cFP^\top)^\top(cFP^\top) = I_n - c^2 P F^\top F
                          P^\top \\
  &= (1-c^2)I_n = 0 \iff c = \pm 1. 
\end{align*}
Hence, when $\tilde S = 0$, the maximum value of $\text{tr}(\tilde B Z)^2$ is $\text{tr}(G)^2$ and this occurs only for $\tilde B = \pm FP^\top$. 
The maximum value of $\text{tr}(\tilde BZ)$ is thereby $\text{tr} (G)$ which occurs when $c = 1$, i.e. for $\tilde B =  FP^\top$. 

When $\tilde S \neq 0$, we need to study the sign of $\text{tr} \left ( PGP^\top \tilde S \right)$. Since $G$ is a diagonal matrix we get
\[
\text{tr} \left ( PGP^\top \tilde S  \right ) = \text{tr} \left ( G P^\top \tilde S P \right ) = \sum_{i=1}^n G_{ii} (P^\top \tilde S P)_{ii}. 
\]
We have assumed $Z$ to have full rank, so all singular values of $Z$ are strictly positive, i.e. $G_{ii} > 0$ for each $i$.  
Let $\tilde S$ have singular value decomposition $\tilde S = WJW^\top$. We then get
\begin{align*}
(P^\top \tilde S P)_{ii} &= \left (P^\top W J W^\top P\right )_{ii} = 
\left (
(W^\top P) ^\top  J W^\top P 
\right )_{ii}\\
&=
\sum_{k=1}^n
J _{kk} \left ( W^\top P \right)^2_{ki}.
\end{align*}
Since we have  assumed  $\tilde S \neq 0$ at least one of the singular values of $\tilde S$ must be strictly positive, i.e. we have at least one $J_{kk} > 0$. 
Without loss of generality we assume in the following that $J_{11} > 0$. 
Since both $P$ and $W$ are orthogonal matrices $W^\top P$ is also orthogonal. 
Thereby there exists at least one index $i$ such that $(W^\top P)_{1i} > 0$. For this value of $i$ we then have
\[
(P^\top \tilde S P)_{ii} \geq J_{11} (W^\top P)^2_{1i} > 0.
\]
Thereby, since $P^\top \tilde S P$ is positive semidefinite,
\[
\text{tr} \left ( P GP^\top \tilde S \right ) \geq G_{ii} (P^\top \tilde S P)_{ii} > 0. 
\]
Thus,
\[
| \text{tr}(\tilde B Z) | \leq \sqrt{ \left ( \text{tr}(G) - \text{tr}(PGP^\top \tilde S) \right) \text{tr} (G) } < \text{tr} (G).
\]
We thereby see that the maximum value of $ \text{tr}(\tilde B Z) $ when $\tilde S \neq 0$ is \emph{smaller} than its maximum value when $\tilde S=0$. 
The maximum value of $\text{tr} (\tilde BZ)$ must therefore occur when $\tilde S=0$ and $\tilde B = FP^\top$, and the proof is complete.

\end{document}